# SamurAI: a Versatile IoT Node with Event-Driven Wake-Up and Embedded ML Acceleration

Ivan Miro-Panades, *Member, IEEE*, Benoit Tain, Jean-Frédéric Christmann, *Member, IEEE*, David Coriat, Romain Lemaire, Clément Jany, Baudouin Martineau, Fabrice Chaix, Guillaume Waltener, Emmanuel Pluchart, Jean-Philippe Noel, Adam Makosiej, Maxime Montoya, Simone Bacles-Min, David Briand, Jean-Marc Philippe, Yvain Thonnart, *Member, IEEE*, Alexandre Valentian, Frédéric Heitzmann, Fabien Clermidy

*Abstract*— **Increased capabilities such as recognition and self-adaptability are now required from IoT applications. While IoT node power consumption is a major concern for these applications, cloud-based processing is becoming unsustainable due to continuous sensor or image data transmission over the wireless network. Thus optimized ML capabilities and data transfers should be integrated in the IoT node. Moreover, IoT applications are torn between sporadic data-logging and energy-hungry data processing (e.g. image classification). Thus, the versatility of the node is key in addressing this wide diversity of energy and processing needs. This paper presents SamurAI, a versatile IoT node bridging this gap in processing and in energy by leveraging two on-chip sub-systems: a low power, clock-less, event-driven Always-Responsive (AR) part and an energy-efficient On-Demand (OD) part. AR contains a 1.7MOPS event-driven, asynchronous Wake-up Controller (WuC) with a 207ns wake-up time optimized for sporadic computing, while OD combines a deep-sleep RISC-V CPU and 1.3TOPS/W Machine Learning (ML) for more complex tasks up to 36GOPS. This architecture partitioning achieves best in class versatility metrics such as peak performance to idle power ratio. On an applicative classification scenario, it demonstrates system power gains, up to 3.5x compared to cloud-based processing, and thus extended battery lifetime.**

## I. Introduction

NOWADAYS, the IoT domain uses traditional cloud computing paradigms to process the collected data [1]. The data is collected by IoT nodes, where it can be filtered, compressed and encrypted. Then, the data is streamed through gateways to the cloud where it is processed [2]. However, as the number of IoT devices grows exponentially, a centralized cloud-based computing will become unsustainable. Moreover, the data processing latency and the unreliability of the network will also limit the scalability of the current approach [3]. The edge-computing paradigm tries to balance the processing of the data by moving it closer to where it is collected [1]. Smart IoT nodes where the data collection and processing are performed locally can address the scalability, latency and unreliability issues of traditional approaches. Thus, machine learning is seen as an enabler for these new class of nodes [4]. However, moving the computation into the node increases the energy spent by the node to process this data while reducing the energy spent streaming the collected data. As these nodes are powered by small batteries, the amount of available energy at the node is limited. Thus, new challenges appear for these nodes. Firstly, energy efficiency is key for these devices, as local data processing has a high energy cost. Secondly, they require more embedded memory than the current IoT nodes in order to process the data locally. Moreover, the node should be versatile enough to efficiently address the high energy difference between idle, data-logging, and local data processing modes.

To address these new challenges, many research efforts are performed to optimize the nodes across multiple domains such as the power consumption of the node itself, the radio communications and the machine learning accelerators.

## II. Context and Motivation

### A. Power consumption

Optimizing the power consumption of the node is primordial as the battery lifetime of the node depends on its average power consumption. Moreover, most of the time the processing elements of the sensor are in sleep mode, during which the power consumption is conditioned by the remaining powered-on parts of the circuit. Thus, many research efforts [1][5][6][7] focus on reducing these energy consumptions by reducing the power supply voltage of the node and by optimizing the deep-sleep mode of the architecture. To support low voltage operations, the memory bitcells are also optimized as in [8][9][10].

However, the deep-sleep mode suffers from long (tens of µs) wake-up time due to extremely low voltage and a reduced clock frequency [6][11]. For some applications, a long wake-up time is not compatible with the wake-up event duration, the bursty nature of the incoming data or the power break-even time [12].

Manuscript received April 20, 2022; revised July 6, 2022.

This work was in part funded by the WAKeMeUP project, which received funding from the Electronic Components and Systems for European Leadership Joint Undertaking in collaboration with the European Union's H2020 Framework Programme (H2020/2014-2020) and National Authorities, under grant agreement no. 783176. It is also part funded by SERENE-IoT (Penta 16004), MACS (FUI20) projects.

I. Miro-Panades, J.-F. Christmann, D. Coriat, R. Lemaire, E. Pluchart, J.-P. Noel, S. Bacles-Min, Y. Thonnart, A. Valentian, F. Clermidy are from CEA, List, University Grenoble Alpes, Grenoble, FR. B. Tain, D. Briand are from CEA, List, University Paris-Saclay, Gif sur Yvette, FR. B. Martineau, F. Chaix, G. Waltener, F. Heitzmann are from CEA, Leti, University Grenoble Alpes, Grenoble, FR. C. Jany was in CEA Grenoble, FR, now with Tihive, Grenoble, FR. A. Makosiej was in CEA Grenoble, FR, now with Xenergic, Lund, Sweden. M. Montoya was in CEA Grenoble, FR, now with NXP Semiconductors Austria GmbH & Co KG. J.-M. Philippe was in CEA Paris-Saclay, FR, now with Thales Research & Technology France. (contact e-mail: ivan.miro-panades@cea.fr).

For these cases, the node cannot be pushed into deep-sleep mode and therefore the benefits of power reduction are lost.

Due to their event-driven behavior, asynchronous circuits are a promising approach to implement fast and reactive wake-up schemes but also environment aware duty-cycling of IoT applications [13]. In highly changing environments and low activity contexts, a clock-less processor implemented in asynchronous logic is a relevant solution which offers reduced power consumption, high robustness to operating conditions and short reactivity [14][15]. As no clock is required to operate, the idle energy is limited to leakage and its performances are automatically adapted to the operating conditions.

### B. *Efficient ML at the IoT edge*

IoT devices at the edge are resource and energy constrained. This advocates for efficient hardware ML accelerators, increased but limited memory resources with reduced power consumption, and optimized ML algorithms well suited for these constrained nodes.

In [16] an overview of the architectures for efficient deep learning on IoT devices is presented. Flexible approaches such as GPUs are too power-hungry to be considered for integrating ML in real IoT devices, while FPGAs are not power-efficient enough. Research focuses on specialized accelerators, especially using dataflow paradigms. Eyeriss [17] is an example of such an accelerator where small processing elements using private buffers are interconnected to implement row-stationary dataflow for efficient convolution operations. Other architectures also use a MAC-array structure, such as the Google Edge TPU [18] architecture. These accelerators focus on convolutions, however complete networks may need more flexibility. This is the domain of heterogeneous accelerators such as the Orlando platform from STMicroelectronics [19], in which a dataflow hardware engine is coupled with SIMD DSP clusters.

To be more efficient, algorithm-architecture matching needs to be improved. In [4] algorithmic and processor techniques are studied to address deep learning in the edge. Quantization and network pruning are popular solutions to reduce the memory footprint of a DNN network [20] with reduced classification accuracy degradation. Network pruning is an efficient way to also reduce the computational cost of the network. However, classical network pruning is not a hardware-friendly approach due to the random location of pruned values. Structured sparsity [21] selectively chooses the pruned locations in order to improve the hardware efficacy of the accelerator.

### C. *Radio communication*

In this context, an always-responsive receiver technology continuously monitoring a radio channel while providing a very low power consumption in the order of microwatts and a radio wakeup time in the range of milliseconds is desirable [22]. Consequently, this wake-up radio (WuR) increases operating life significantly compared to a classic radio receiver. The WuR should detect ISM radio signals, typically the 433 MHz, 868/915 MHz or 2.4 GHz frequency bands.

### D. *L-IoT Approach and SamurAI proposal*

IoT nodes must be versatile to address a wide diversity of applications. However, conventional systems do not have enough versatility to address these applications. The concept of L-IoT was introduced in [23] and [24] where a dual sub-subsystem is used: an Always Responsive (AR) part and an On-Demand (OD) part. The AR part is based on low-power wake-up modules able to handle system events while the OD part is based on high performance and energy-efficient modules, which are most of the time powered off. Moreover, by using asynchronous logic technology on the AR part, the system is able to quickly wake-up (e.g. much faster than one period of the conventional 32 kHz crystal clock.) without using a running clock while its power consumption remains low. The combination of these techniques aims to build a versatile node with fast wake-up time and high energy-efficiency.

Thanks to this dual system, the lesser computationally demanding applications can be managed with only the AR part while higher computationally demanding applications can be managed with the combination of both sub-systems (Fig. 1). For example, in a smart camera application, the AR part can be used to trigger the capture and classification of an image by the OD part only when multiple sensors (e.g. sound, pyroelectric) detect simultaneously. As capturing and classifying an image is an energy-demanding task, the AR part is used here to filter the number of times the task is executed based on sensors correlation and on previous classification results. Finally, as the AR part is based on a programmable core architecture, it can adaptively configure OD triggering conditions to limit the false detections.

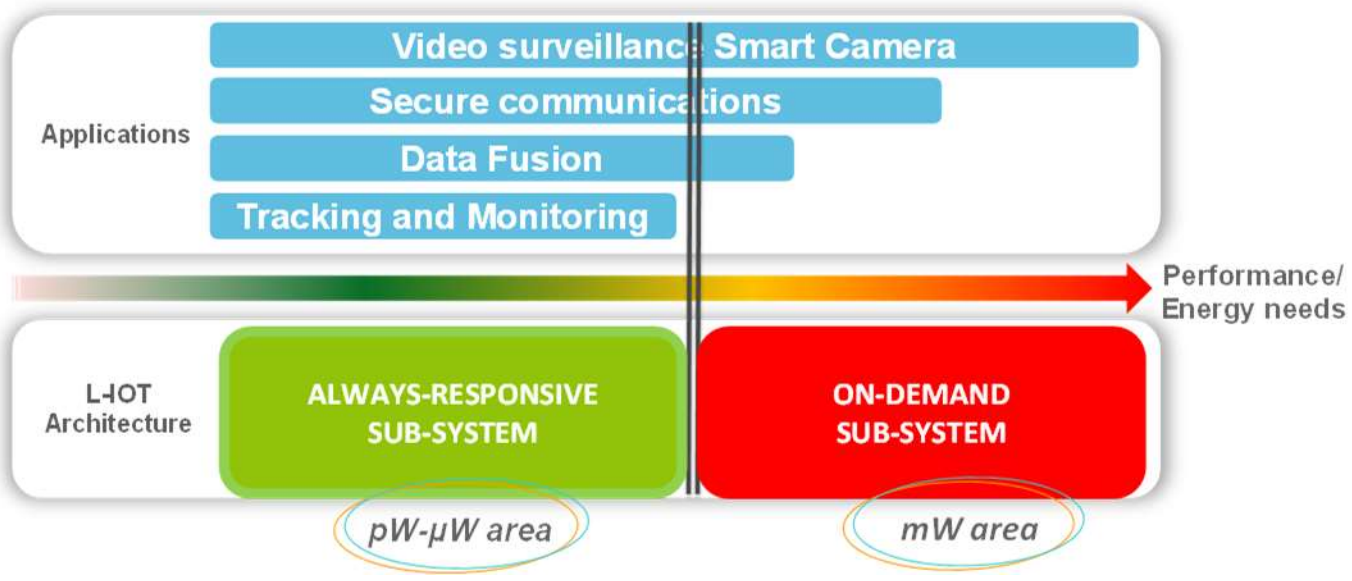

Fig. 1. L-IoT approach.

In [25], an asynchronous circuit design methodology is described for ultra-low power systems. The authors evaluate the gains of an asynchronous wake-up controller on the always-responsive sub-system. Thanks to the fast wake-up time and the low-power consumption during idle states, the authors claim a power gain of up to 70%. Authors in [26] have modeled a wireless sensor network system based on this approach using 28nm FDSOI technology. By using an asynchronous always-responsive sub-system, they claim power gains ranging from 14.5% to 76% for different simulated application targets.

In this context, we propose SamurAI a smart IoT node pushing further the concept of AR and OD part with the integration of WuC, WuR, RISC-V core and PNeuro, an energy-efficient ML accelerator. Specifically, we propose the TP-SRAM, a dedicated two-port SRAM architecture, which allows both asynchronous access at low voltage and high intensity access from RISC-V core or accelerators, while supporting an efficient power gating of OD part. We demonstrate concurrent read/write accesses of AR and OD parts with 15.5ns asynchronous wake-up/sleep modes. Besides, we design a specific instantiation of PNeuro accelerator dedicated to IoT ML operation, such as keyword spotting and image recognition. We demonstrate up to 1.3TOPS/W and 36GOPS computing capability on PNeuro. Conversely, the efficient AR/OD decoupling allows very low power continuous sensing operation at 6.4µW, with 207ns wake-up time for more complex data processing, making SamurAI an energy-efficient versatile IoT node suitable to diverse sensing, recognition and communication applications.

We recently presented an overview and main results of the SamurAI architecture [27]. In this paper, we develop this presentation with additional micro-architecture details and measurements. The outline of the paper is as follows: section III introduces the SamurAI architecture and its power domains. Section IV presents the Always-Responsive sub-system of the architecture while section V describes the On-Demand sub-system part. Finally, section VI and VII present the final circuit results, and application scenario and conclusion.

## III. SamurAI Architecture

An event-driven IoT node is a way to reduce the power consumption of sporadic computing node. SamurAI leverages the L-IoT approach to build a smart node to address the challenge of application versatility. Following the L-IoT approach, it combines a dual-system with a machine learning (ML) accelerator.

### A. *Functional partitioning between event-driven AR and compute-intensive OD subsystems*

Figure 2 presents the SamurAI system architecture. The AR part is composed of an event-driven asynchronous Wake-up Controller (WuC), a Wake-up Radio (WuR) and a Digital Baseband (DBB) demodulator module. The OD part is articulated around a RISC-V core and a ML accelerator (PNeuro). The WuC is the master of the system when the OD part is OFF. Thus, limiting the power consumption to only the AR part when no task is running. If needed, it can wake-up the OD part and define the tasks to be executed by the RISC-V core and PNeuro. Between the AR and OD parts, a low-power two-port SRAM (TP-SRAM) memory is used. It is the WuC's main memory and it is also used for data exchange between the AR and OD parts. This memory must be low leakage with a low operation voltage, when only AR is powered, and high data throughput when both parts are powered to reduce data exchange latency. In terms of radio communication, the WuR and DBB are used to receive small radio messages on IoT bands to control the node with a reduced power budget.

The memory subsystem on RISC-V side is composed of Tightly Coupled Data and Program Memories (TCDM and TCPM) and an external non-volatile memory (FeRAM). Thus, the core sees 192KB of SRAM and 512KB of NVM with in-place execution. The core communicates through an AHB bus with PNeuro, TP-SRAM and an APB bridge to peripherals. The latter are: 1) Crypto IP to encrypt and decrypt messages, 2) Timers, 3) Adaptive Voltage Scaling (AVS) controller to track the Vmin of the circuit, 4) interrupt (IT) controller for RISC-V and WuC events, 5) synchronization (Locks) controller, 6) SPI master, 7) GPIO controller, 8) I2C controller, 9) UART controller. WuC can also communicate with these peripherals through an APB bridge between the AR and OD parts. Moreover, WuC has a dedicated GPIO interface for sensor connectivity and to wake-up on sensor events.

The node is equipped with an SPI slave and JTAG interface to program and debug the system. Moreover, timers and performance monitors are used for system ticks and to measure the circuit performance.

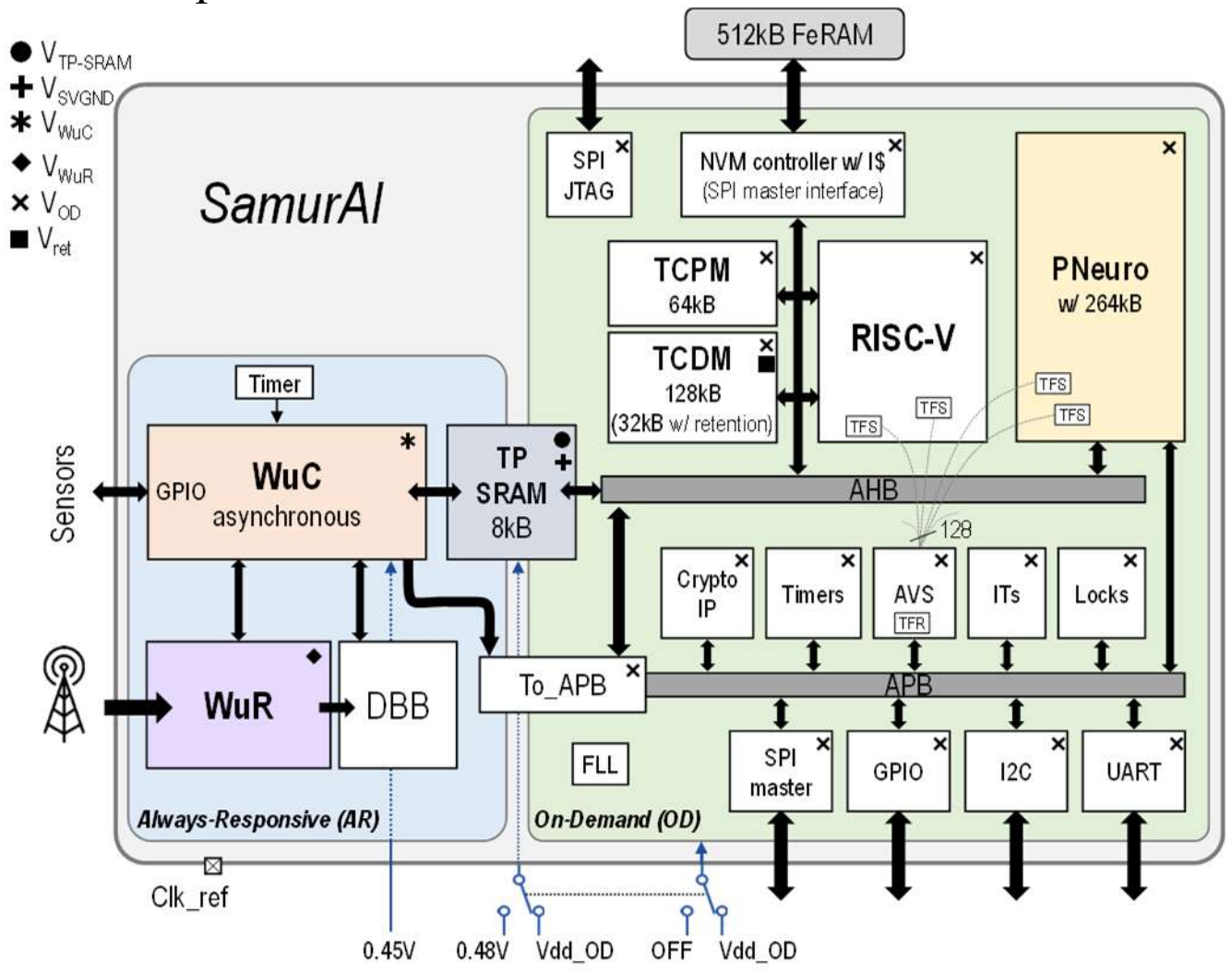

Fig. 2. SamurAI system architecture, with Always-Responsive and On-Demand sub-systems and associated power domains.

### B. *Clocking and power domains*

In the AR part, the WuC is implemented with Quasi Delay Insensitive (QDI) [28] fully-asynchronous logic, the WuR uses a switchable free-running VCO, while the OD part can be clocked either by an external clock (clk_ref) or by an internal FLL clock generator (based on [29]). Moreover, this FLL has a programmable free-running capability to generate a clock when the external clock is disabled. The TP-SRAM offers a dual clocking mode, with fully-asynchronous operation when used only by the AR part and a synchronous mode when used with the OD part.

The node has five power modes (Table I) where the voltage of the AR, TP-SRAM and OD parts are generated externally and controlled by the WuC through external power switches.

Table I: SamurAI power modes

| | Always-Responsive (AR) | | | TP-SRAM | | On-Demand (OD) | | |
|---|---|---|---|---|---|---|---|---|
| Power mode | Voltage (V) | WuC state | WuR State | Voltage (V) | State | Voltage (V) | RISC-V Freq (MHz) | Periph Freq (MHz) |
| IDLE | 0.45 | Sleep | OFF | 0.48 | Sleep | OFF | - | - |
| WuC only | 0.45 | Run | OFF | 0.48 | Run | OFF | - | - |
| WuC+WuR | 0.45 | Run | ON | 0.48 | Run | OFF | - | - |
| WuC+Periph. (cpu sleep) | 0.45 | Run | ON/OFF | 0.48 | Run | 0.48 | Gated | 10 |
| CPU running | 0.45 | Run/Sleep | ON/OFF | 0.48 – 0.9 | Run | 0.48 – 0.9 | 1 – 350 | 1 – 350 |

In *IDLE* state, TP-SRAM is in sleep mode (retentive mode with power gating on the periphery) and WuC is waiting for an event to wake-up. These events can come from sensors (GPIOs), DBB, timer or the OD part. When an event happens, WuC wakes up TP-SRAM and executes the code from TP-SRAM to manage the event (*WuC only* power mode). At the end of the execution, the WuC can change the power mode to 1) use the WuR with *WuC+WuR* mode, 2) use the OD peripherals with *WuC+Peripherals* mode, 3) execute tasks on RISC-V core with *CPU running* mode, or 4) go back to sleep with *IDLE* mode.

TP-SRAM is kept at low voltage when only the AR part is ON to reduce the leakage. However, its voltage is increased and linked with the OD part, when the latter is powered on (*CPU running* mode) to maximize the data exchange throughput between AR an OD parts.

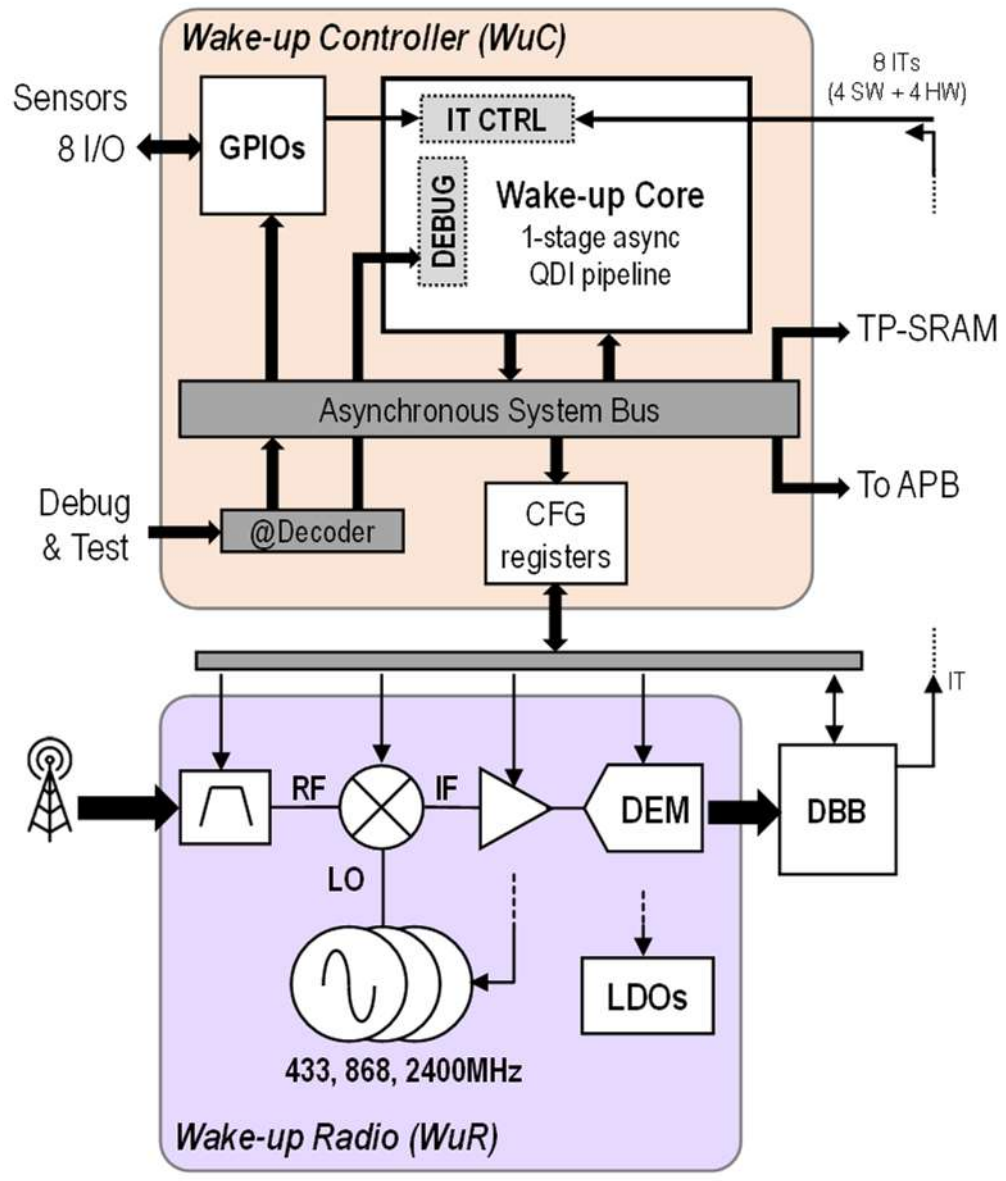

Fig. 3. Wake-up Controller and Radio architecture details.

## IV. Always-Responsive (AR) Sub-System

### A. *Wake-up Controller*

The WuC (Fig. 3) is a clock-less 32b MCU with a 16b RISC ISA, based on [15] and [30]. The WuC is the master of the AR sub-system. It manages the sensor data, the wake-up events and controls the power modes of the circuit. Thanks to asynchronous Quasi Delay Insensitive (QDI) design techniques [28], core performance is automatically adapted to the PVT variations without requiring any timing assumption. Tiempo [31] Asynchronous Circuit Compiler (ACC) has been used for the logical synthesis.

Fig. 3 shows the WuC architecture, where Wake-up Core is the computing core, a single issue, single pipeline core. The Asynchronous System Bus is the local interconnect which interfaces the processor core, the GPIO controller, the configuration (CFG) registers, the TP-SRAM, and the OD APB interface. The TP-SRAM contains the program and data for the core and is accessed using an asynchronous handshaking protocol (see Section IV.C) to avoid using a clock frequency when only WuC uses the memory. The APB interface to the peripherals uses a synchronous 4-phase handshake interface to exchange the data from the asynchronous domain to the synchronous OD domain. The attached GPIO controller manages 8 input/output circuit pads on which sensors can be connected. These signals can be used as wake-up sources. An interrupt controller (IT) selects the signals (8 GPIOs and 8 internal sources) and triggering conditions to wake-up the WuC. Among the internal interrupt sources, four are hardware (HW) interrupts (one from DBB and three from the OD sub-system), and four are software (SW) interrupts for inter-task synchronization, and debug & test functionality. Finally, the CFG registers are used to configure the circuit including the OD sub-system, WuR and DBB parameters.

The core uses a run-to-completion scheduling model to simplify its architecture. By default, the core waits for an incoming interrupt (8 from GPIO and 8 from internal sources). Once an interrupt arrives, a dedicated routine is executed until completion. If new interrupts have arrived in between, their dedicated routines are then executed. Otherwise, the core enters into the wait for event mode (IDLE). In an asynchronous logic scheme, this mode implies zero signal toggling, hence zero dynamic power consumption.

### B. *Wake-up Radio and DBB*

The goal of the Wake-Up Receiver is to sense the channel instead of main radio receiver for a lower power consumption. To be effective, the power consumption of this wake-up device must be in the range of a decade below the main receiver. In a multiple sensors IoT scenario, multiple frequency operation is essential. The proposed circuit is derivative from [32] and is based on a 2-path passive mixers first topology (as in [33]) that allows a wide band 50 Ohms input matching from 400MHz to 2.5GHz (Fig. 3). This large bandwidth has been chosen to be compatible with ISM radio band around 433MHz, 868MHz, 902MHz and 2.4GHz frequencies. The frequency band selection is performed thanks to three independent current-starved ring oscillators. The inaccuracy of the local oscillator generates an uncertain IF signal when mixing with the RF input desired signal. At the IF-band, a specific capacitor is used at the output of the passive mixer. This effect creates a band-pass like input impedance response, whose bandwidth gets reduced as the IF capacitor is increased. IF signal is then amplified through a 3-stage IF chain with programmable gain and bandwidth. Finally, the signal is demodulated by self-mixing through an envelope detector combined with a limiting amplifier allowing the full SOOK demodulation.

The digital baseband (DBB) module decodes small radio messages composed of 8b identifier to selectively wake-up the WuC and a 32b message payload for application specific purposes. It has been designed in a generic manner, leveraging symbol width and useful data delay within the symbol as dynamic parameters, and may thus be configured to handle any kind of OOK modulation. Once the identifier has been detected and the payload has been received, an interrupt is sent to the WuC, triggering the execution of a routine dedicated to RF wake-up events.

## C. TP-SRAM

The TP-SRAM is at the intersection of the AR and OD sub-systems. Its role is to serve as main memory for the WuC and to exchange data between the two sub-systems. As it contains the WuC's program and data, it is always powered, thus its power consumption must be optimized both in idle and in active mode. To leverage the benefits of the asynchronous architecture and its fast reaction time, TP-SRAM has been designed as a clock-less memory with automatic power-up/down capabilities [34]. A low-voltage scheme is used, allowing for a fast reaction time with read/write capabilities.

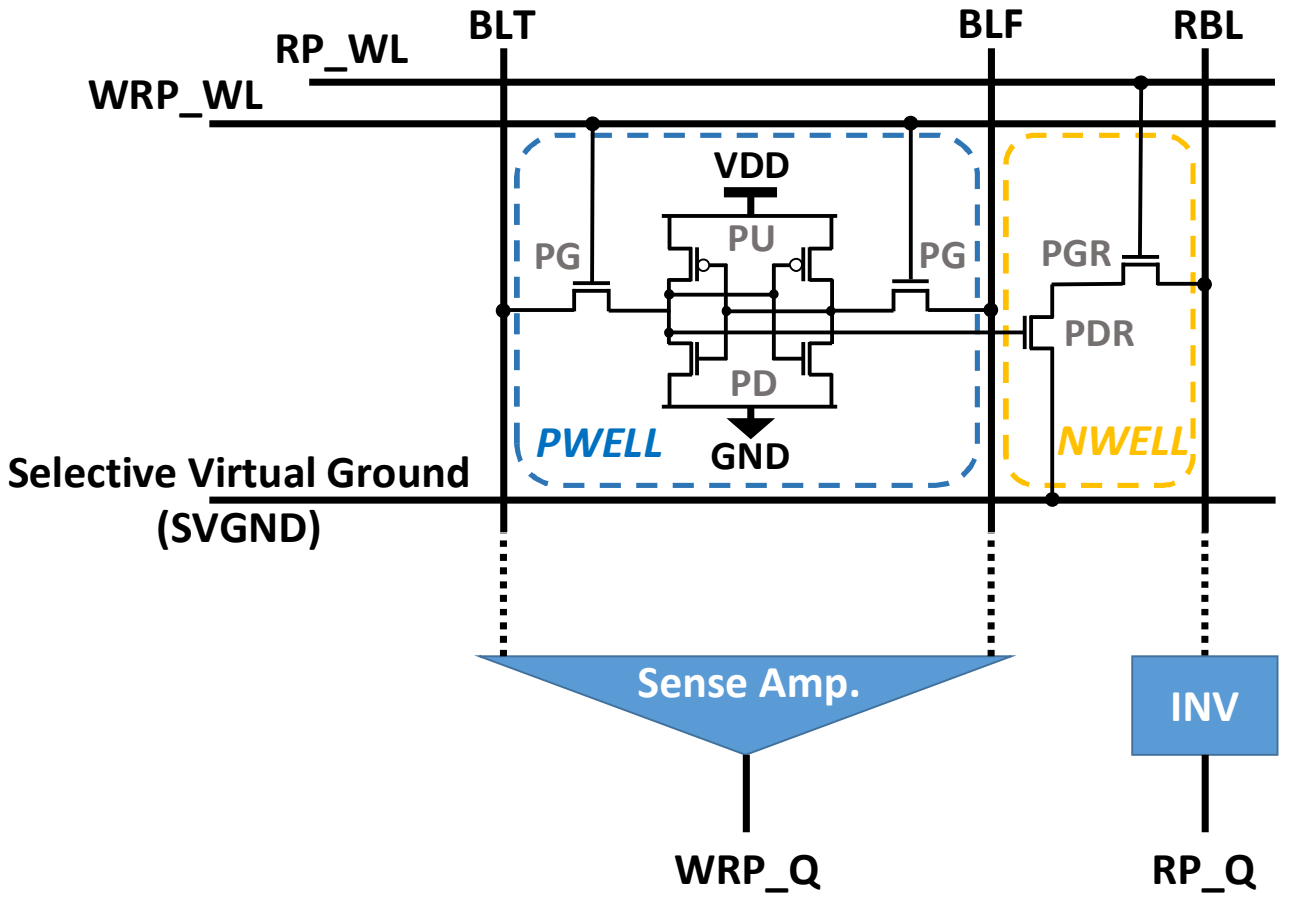

Fig. 4. TP-SRAM bit cell architecture.

To support low voltage read/write operations, a foundry two-port bit cell (six-transistor write/read port and two-transistor read port) has been used, as shown in Fig. 4.

The write/read port and the read port are controlled by two different word lines (WRP_WL and RP_WL). Thus, it supports two-port operations, a write-read port (WRP) using the six-transistor cell, and a read port (RP) using the two-transistor read port. This feature allows an increased throughput when the WuC and RISC-V share data. However, this concurrent operation is not possible at low voltage as WRP can perform write but not read operations due the minimum offset required by the sense amplifier. Nevertheless, thanks to full swing reading mode, the read port (RP) can be used at low voltages to read the memory content.

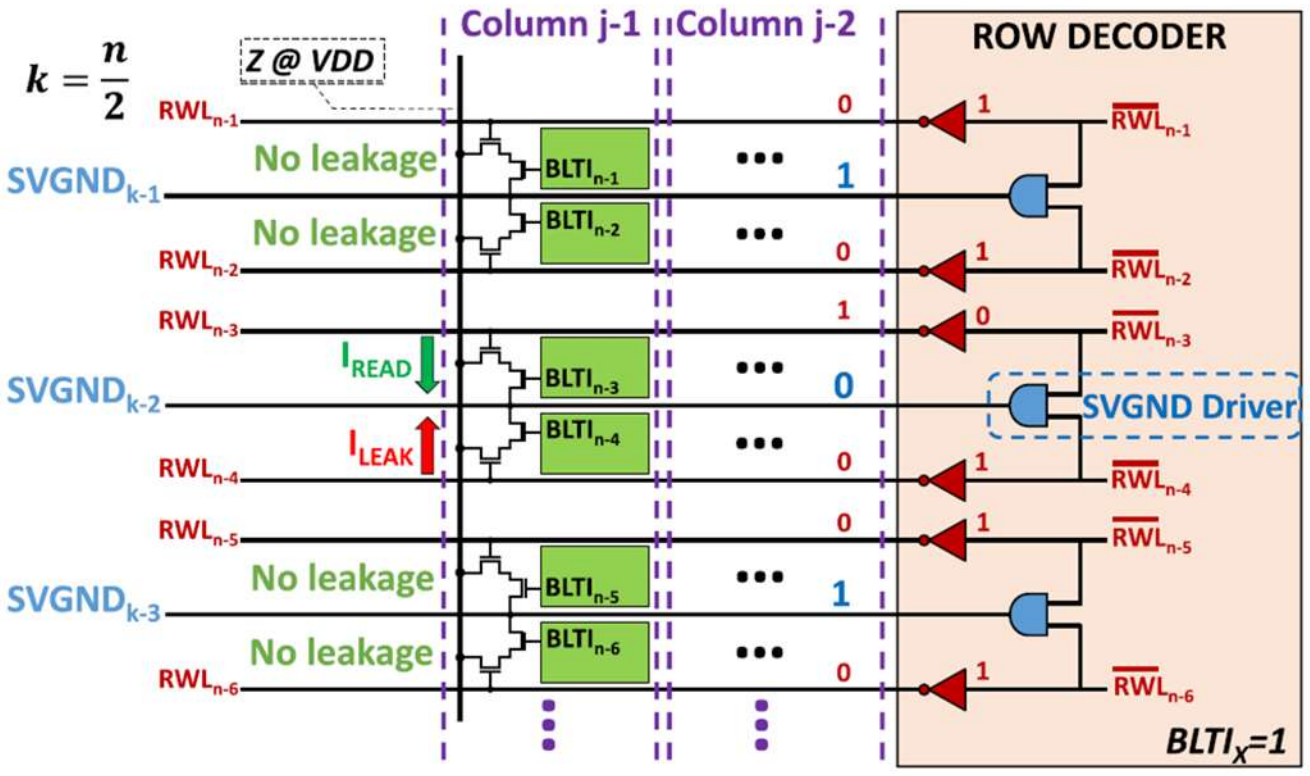

Fig. 5. TP-SRAM read port principle.

To minimize the read port leakage on the read bit line, the virtual ground of every pair of read port word-lines is horizontally controlled between GND and VDD at the row decoder level (Fig. 5). When the word-lines are not selected, SVGND is tied to VDD, the same potential as the idle precharge voltage of the RBL, leading to the memory leakage reduction. When a word-line is selected by enabling RWL, the read margin (difference between logic '0' and '1') is maximized by connecting SVGND to GND to perform a proper read.

The TP-SRAM uses handshake asynchronous protocol interfaces (Fig. 6) to interface with the clock-less WuC for performing read/write operations and to manage its internal power state (sleep mode where the periphery is powered off). There are three handshake interfaces 1) SLEEP_REQ and SLEEP_ACK to manage the sleep mode, 2) WRP_CK and WRP_RDY to manage the start/end of an operation on WRP, and 3) RP_CK and RP_RDY to manage the start/end of an operation on RP. These signals are described hereafter.

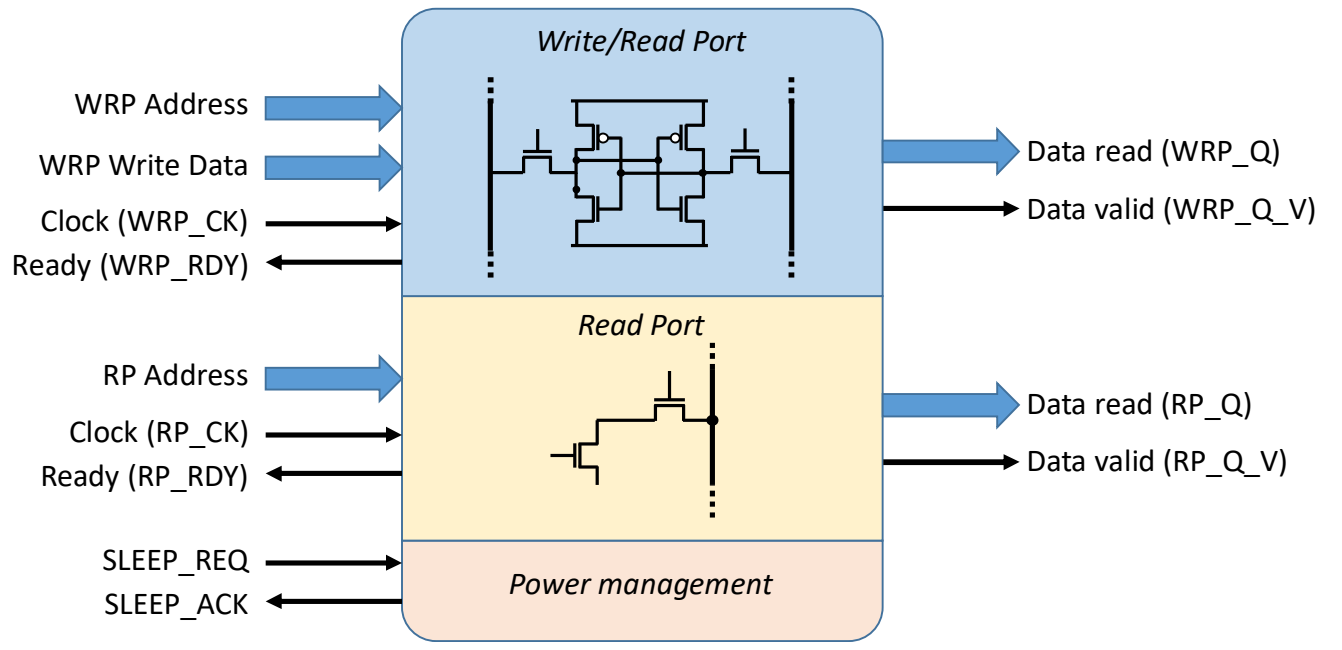

Fig. 6. TP-SRAM interfaces.

Fig. 7 shows a chronogram of the memory starting from sleep mode, waking up, reading two data on RP and entering sleep mode again. When the WuC lowers the SLEEP_REQ signal, the memory exits sleep mode. The TP-SRAM periphery is powered on, and as soon as the memory is ready to perform a read/write operation, it raises the SLEEP_ACK signal. Then the WuC can start an operation to the memory. When the WuC wants to perform a read operation on the RP, it presents the read address and then raises the RP_CK signal. The RP_CK (and WRP_CK) behaves as a clock signal for the memory. The memory, 1) lowers the RP_Q_V meaning that the output data (RP_Q) is no more valid, 2) registers the read address, 3) starts the internal read operation, and 4) lowers the RP_RDY, acknowledging the requested operation to the WuC. When the WuC sees RP_RDY at low, it lowers the RP_CK signal and waits for the requested data. Once the memory has performed the read operation, the data is available at the RP_Q interface and then RP_Q_V is risen to signal the validity of the data. After that, the WuC captures the data. Finally, the memory raises RP_RDY when it can accept a new operation allowing the WuC to start a new request. The same protocol is used on the WRP where read and write operations are possible. In the case of write operations, WRP_Q and WRP_Q_V are not used.

To enter sleep mode, the WuC raises the SLEEP_REQ signal and the memory in turn lowers SLEEP_ACK and powers off its periphery to lower the leakage.

The integration of TP-SRAM into the SamurAI architecture is depicted in Fig. 8 where three power domains are also shown (pd_wuc, pd_tpsram and pd_on_demand).

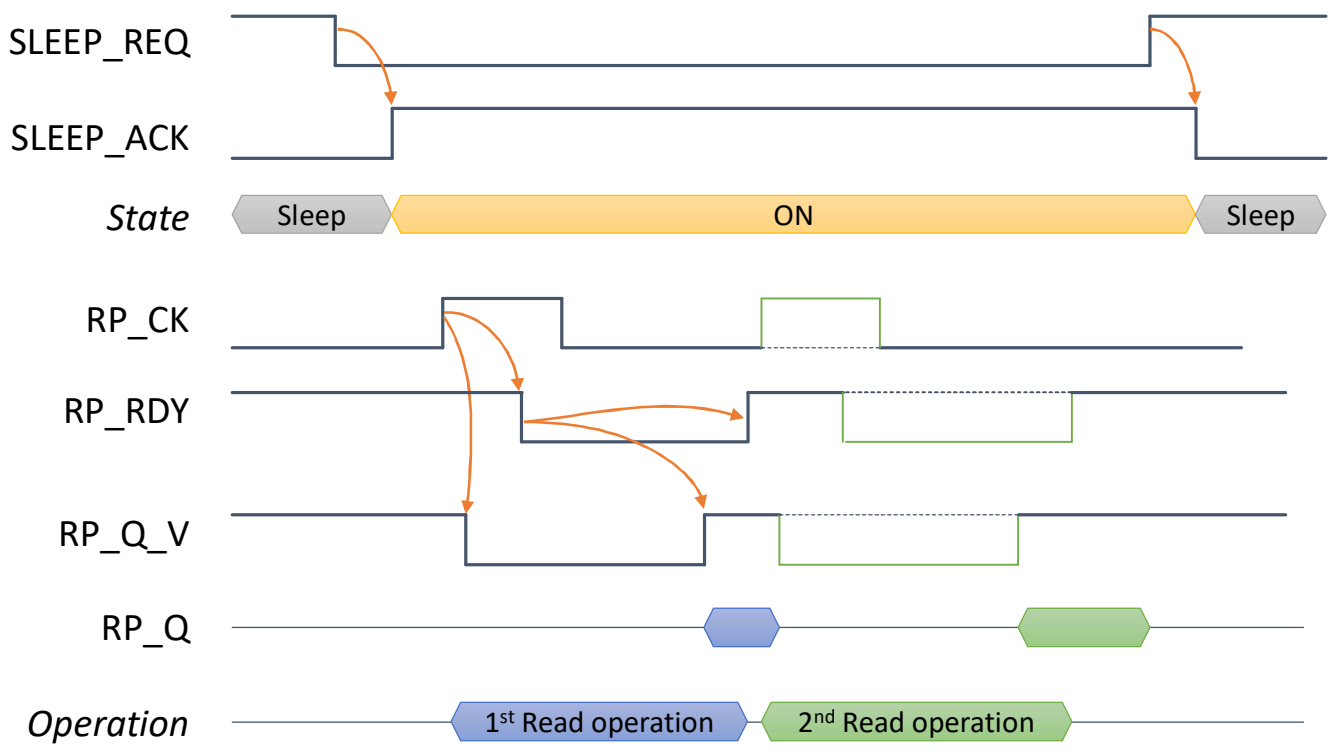

Fig. 7. TP-SRAM chronogram.

The WuC's read and write operations are redirected into two distinct interfaces. Read operations are directly connected to the RP of the TP-SRAM while the write operations are combined with the AHB request into the WRP of the TP-SRAM. Thus, the WuC has a dedicated read interface for fetching instructions and reading data while it interleaves the write operations with the OD sub-system. Therefore, WuC can simultaneously fetch a new instruction while writing data into the memory. When the OD sub-system is powered OFF, the WRP is directly connected to WuC, allowing fast asynchronous operations. However, when the OD sub-system is used, a synchronous round robin arbiter performs the allocation of the WRP to either WuC or AHB interfaces. In this case, the WRP is clocked by the OD clock (clk_od) in order to maximize the throughput for AHB requests while the asynchronous WuC's write operations are converted into synchronous requests with a 4-phase protocol converter.

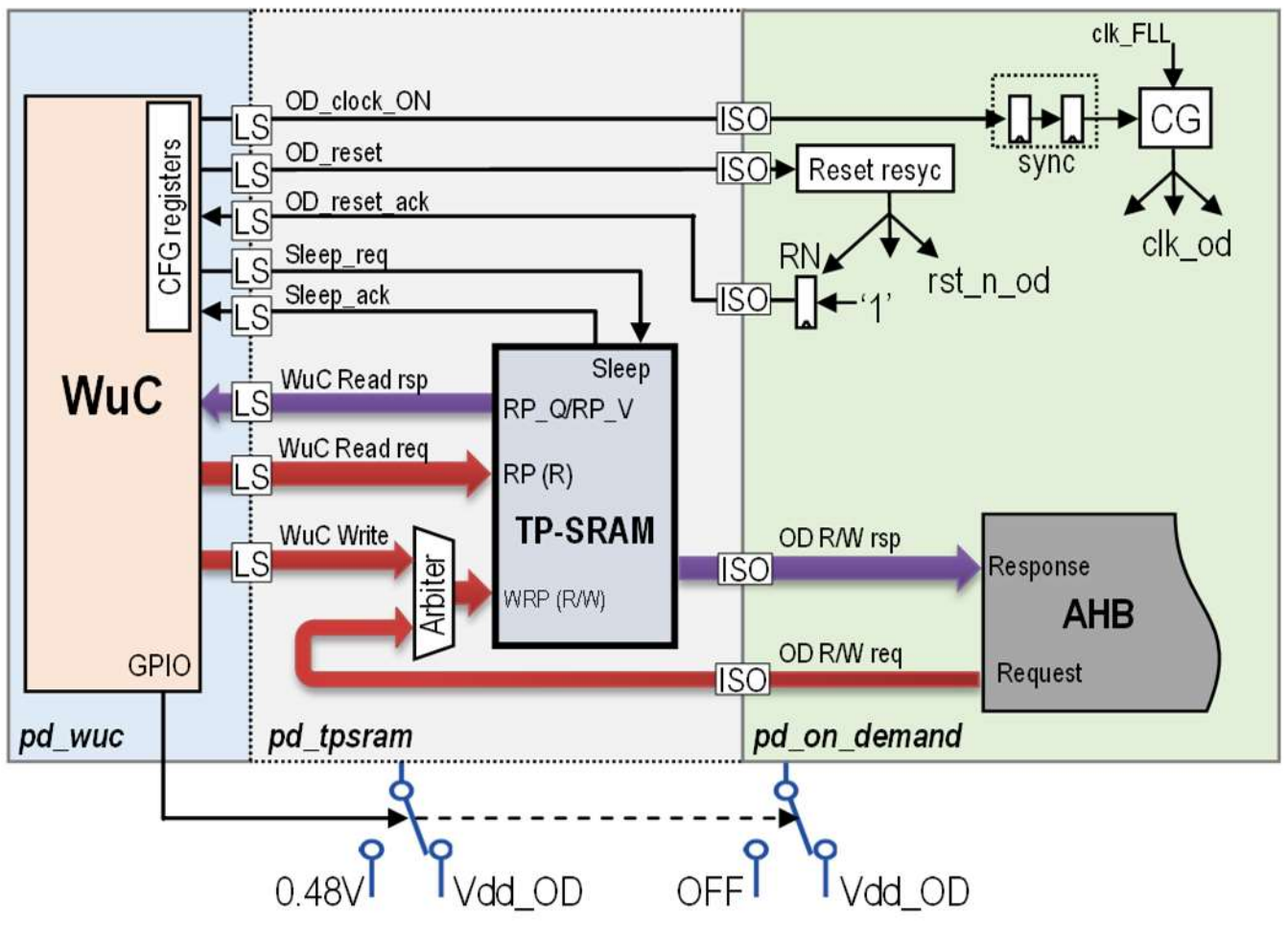

Fig. 8. TP-SRAM interfaces and OD control registers.

To simplify the power and clocking strategy, the pd_tpsram and pd_on_demand power domains are linked the when OD is power on. Thus, only isolation cells (ISO) are used between this two power domains (Fig. 8). However, level shifters (LS) are inserted between pd_wuc and pd_tpsram power domains to allow for independent voltage modifications.

The clock frequency (FLL) generation, the clock-gating and reset of the OD domain are managed through WuC's configuration registers. A handshake protocol on the reset signal (OD_reset and OD_reset_ack) guarantees that the OD domain has ended the reset phase before interacting with it, making the system robust to any clock frequency and PVT variation. This mechanism is also used to switch the arbitration policy of WRP port.

## V. On-Demand (OD) Sub-System

### A. *RISC-V and Memory Sub-System*

The main core on the OD (Fig. 2) sub-system is the RISC-V (RV32IMCXpulp) core from Pulp project [35]. It is a single-issue, in-order core with 4-stage pipeline. The core has post-incrementing load and store operations, multiply-accumulate operations and hardware loops to speed up the execution of the code. As the RISC-V core is most of the time off, the WuC behaves as the master of the node. The fetch enable and the boot address of RISC-V core is controlled by the WuC. Thus, the WuC can power on the OD sub-system and configure the task to start on RISC-V core by changing its boot address to a different NVM location.

The memory sub-system is composed of 128kB TCDM, 64kB TCPM and of 512kB external non-volatile memory (NVM). The tight-coupled memories are directly visible by the core and mapped on the global memory space. 32kB of TCDM memory uses a retention SRAM to keep the RISC-V task context when the OD sub-system is powered OFF.

The external NVM is managed by the *NVM controller*, which is composed of an instruction cache controller, a FeRAM controller, and an SPI master interface. The RISC-V instruction fetches are routed to an instruction cache controller while the data request are directly routed to the FeRAM controller. This controller arbitrates between instruction and data requests and manages the SPI link with the external FeRAM.

The instruction cache controller uses a direct map architecture with 4 sets and 8 words of 32b allowing in-place execution. Thanks to this cache, the SPI efficiency reaches 91%, with 24b for control and 256b for payload.

### B. *PNeuro Accelerator*

PNeuro [36] is a flexible and low-power accelerator dedicated to speed up the inference part of Deep Neural Network (DNN) algorithms. Its architecture is presented in Fig. 9. PNeuro is based on a set of computing clusters interconnected with AHB and APB busses. These computing clusters are composed of a cluster controller which fetches and decodes instructions from the 4kB program memory. The decoded instructions are then sent to a set of Neural Compute Blocks (NCB), which are responsible for the different computations.

Each NCB is built around a 32kB multi-banked SRAM, enabling parallel access to different data commonly found in DNN, such as the weights and data but also intermediate data required by the computation process. These SRAM are driven by different address generators that automatically implement address generations for common DNN operations (such as convolution or fully-connected layers). Computations are performed at the NCB level by processing elements (PE). There are 8 PEs per NCB in the presented configuration. All PEs of a cluster form an SIMD (Single Instruction Multiple Data) processor. Local control implements traditional flags and guards management. Each NCB also features a routing unit, which is responsible for transmitting the data flows between the SRAM banks and the PEs. Besides data flow transformation, such as broadcasting, this unit can also perform data injection for automatic padding management.

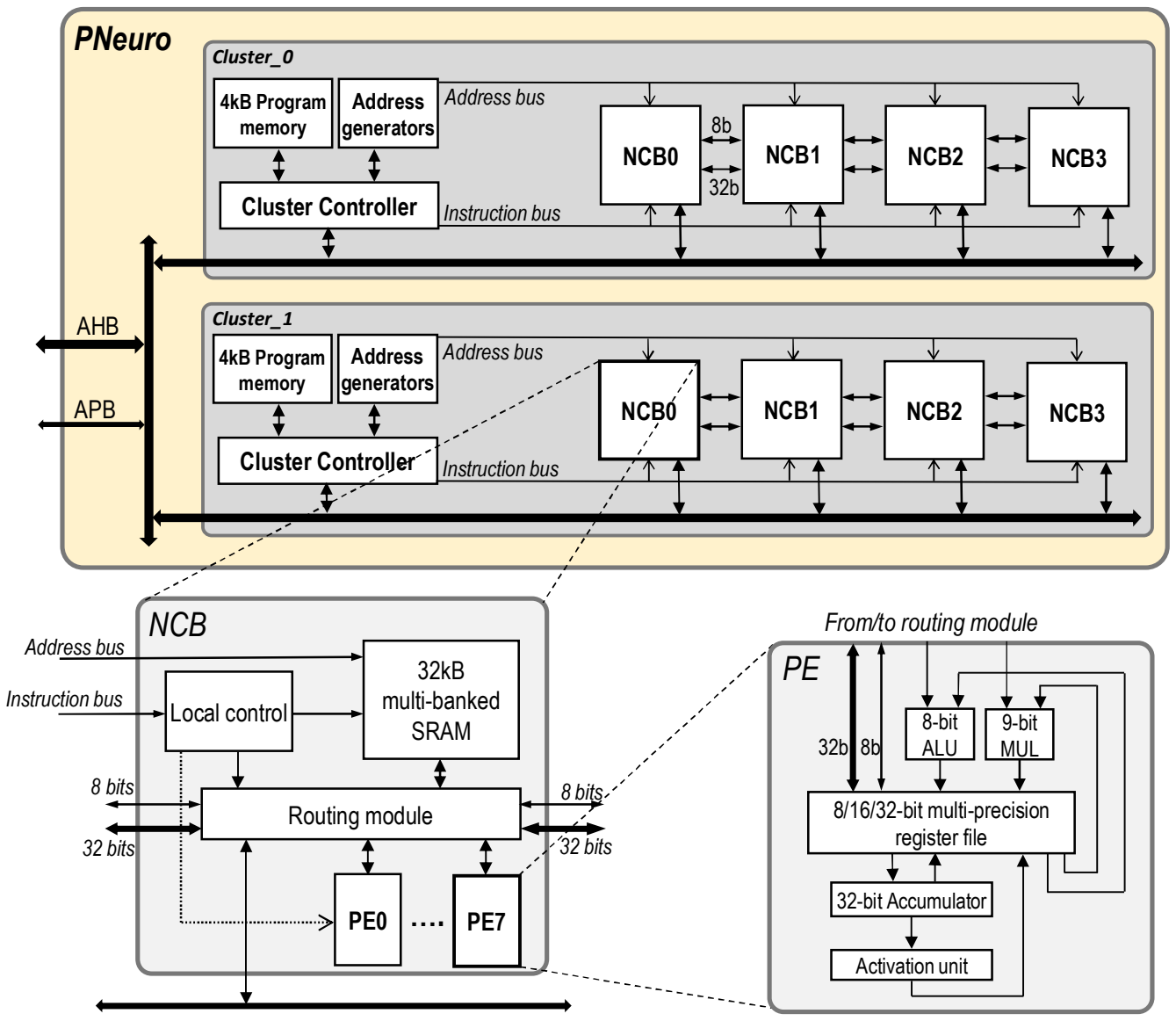

Fig. 9. Two-cluster PNeuro accelerator with 64 PEs.

Each PE provides the necessary operators to perform DNN-related computations. PNeuro computes 8-bit data natively; hence, it provides an 8-bit ALU. In order to perform full precision computation of image data and signed-weights, 9-bit multipliers are used as a first stage for the multiply-accumulate (MAC) operator. An intermediate multi-precision register file enables to implement complex dataflows or signal processing operations in which only multiplications are needed (without accumulation). A 32-bit accumulator implements the second stage of the global MAC operator in the PE. Finally, an activation unit implements non-linear operations such as the linear rectification. Inter-PE communication is ensured by two interconnects (8-bit and 32-bit respectively), which perform direct neighbor connections through the entire cluster.

SamurAI features a PNeuro accelerator with two clusters of 4 NCB, each comprising 8 PE. This configuration can thus output 64 MAC operations per clock cycle. PNeuro is a programmable architecture with two types of instructions: control instructions directly perform operations on the cluster controller and the address generators, while compute instructions relate to operations in the processing elements. Synchronization with the RISC-V core is performed using memory-mapped registers enabling the host processor to trigger execution of the clusters individually. PNeuro cluster controllers can send interrupts to the RISC-V core for signaling events related to computations (end of computation, start of data transfer, etc.). The PNeuro memory space is entirely memory-mapped in the system, enabling the RISC-V to transfer data and DNN parameters while retrieving results and intermediate data.

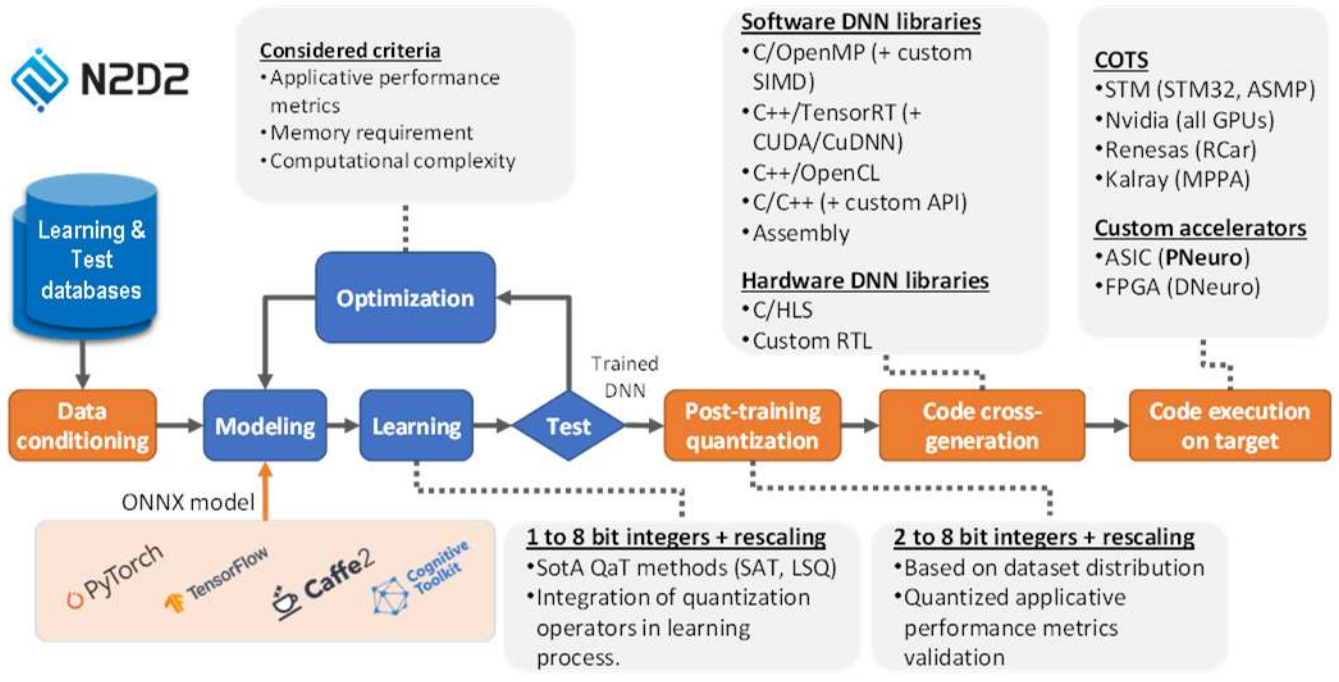

Fig. 10. N2D2 framework.

Besides the hardware architecture, a software platform based on the export module of N2D2 [37] was developed. N2D2 (for Neural Network Design and Deployment) is a deep learning framework tailored for quantization aware training and optimized code generation for various embedded hardware targets (Fig. 10). The framework generates code to multiple targets including CPU, GPU and ASIC (PNeuro). For the latter, it generated the code for the PNeuro and its host, the RISC-V core. N2D2 provides a complete design environment for a wide range of quantization modes to achieve the best performances. N2D2 implements the latest quantization methods including SAT [38] and LSQ [39] methods achieving an accuracy of 72.60% on MobileNet-v1 with 8-bit quantization for weights and activations. Thanks to N2D2 a wide range of neural networks can be easily deployed on PNeuro.

### C. System Peripherals

The IoT ecosystems are evolving with new applications, new standards and new security threats. Therefore, versatile solutions for IoT security are required to adapt to these factors. SamurAI embeds a Crypto IP module (based on [40] and [41]) with three standardized algorithms (Table II) which can be used for ultra-low-power IoT devices. The block cipher *AES* focuses on the security feature which has high levels of security; while the lightweight block cipher *Present* and stream cipher *Trivium* focus on the effectiveness of the algorithms not only in security feature but also in terms of performance, power budget, and throughput.

TABLE II: IMPLEMENTED CRYPTOGRAPHIC ALGORITHMS

| Algorithm | Type | Key size (bits) | Block size (bits) |
|---|---|---|---|
| AES | Block cipher | 128, 192 and 256 | 128 |
| Present | Block cipher | 80 | 64 |
| Trivium | Stream cipher | 80 | NA |

SamurAI is equipped with an adaptive voltage scaling (AVS) (as in [42] [43]) controller to estimate and track the minimal operating voltage (Vmin) of the circuit for a particular target frequency. AVS manages 128 timing sensors (TFS) to estimate

Vmin and a programmable replica path (TFR) to track it during runtime. The Vmin estimation process is performed in the field, where the circuit is configured to execute a functional test pattern in a loop manner as the power supply is reduced. The TFS sensors trigger when the slack time of a path is lower than a threshold. Compared to canary flip-flops, TFS triggers much earlier. Thus, the circuit is never pushed into failure. To estimate the Vmin of the circuit, the annotated TFS triggering voltages during the functional test are combined with a precomputed equation, leading to as small as a 2% voltage error in the estimation voltage [42] [43]. The parameters of this equation are obtained by measuring Vmin on corner samples of a subset of circuits and correlating TFS triggering information with it. In SamurAI, the algorithm and the equation are computed by the RISC-V core to estimate the Vmin. The estimated Vmin is then programmed into the TFR in order to track the Vmin during circuit runtime. Running at the estimated Vmin instead of SignOff voltage reduces the power consumption between 19% and 39%, depending on the application scenario.

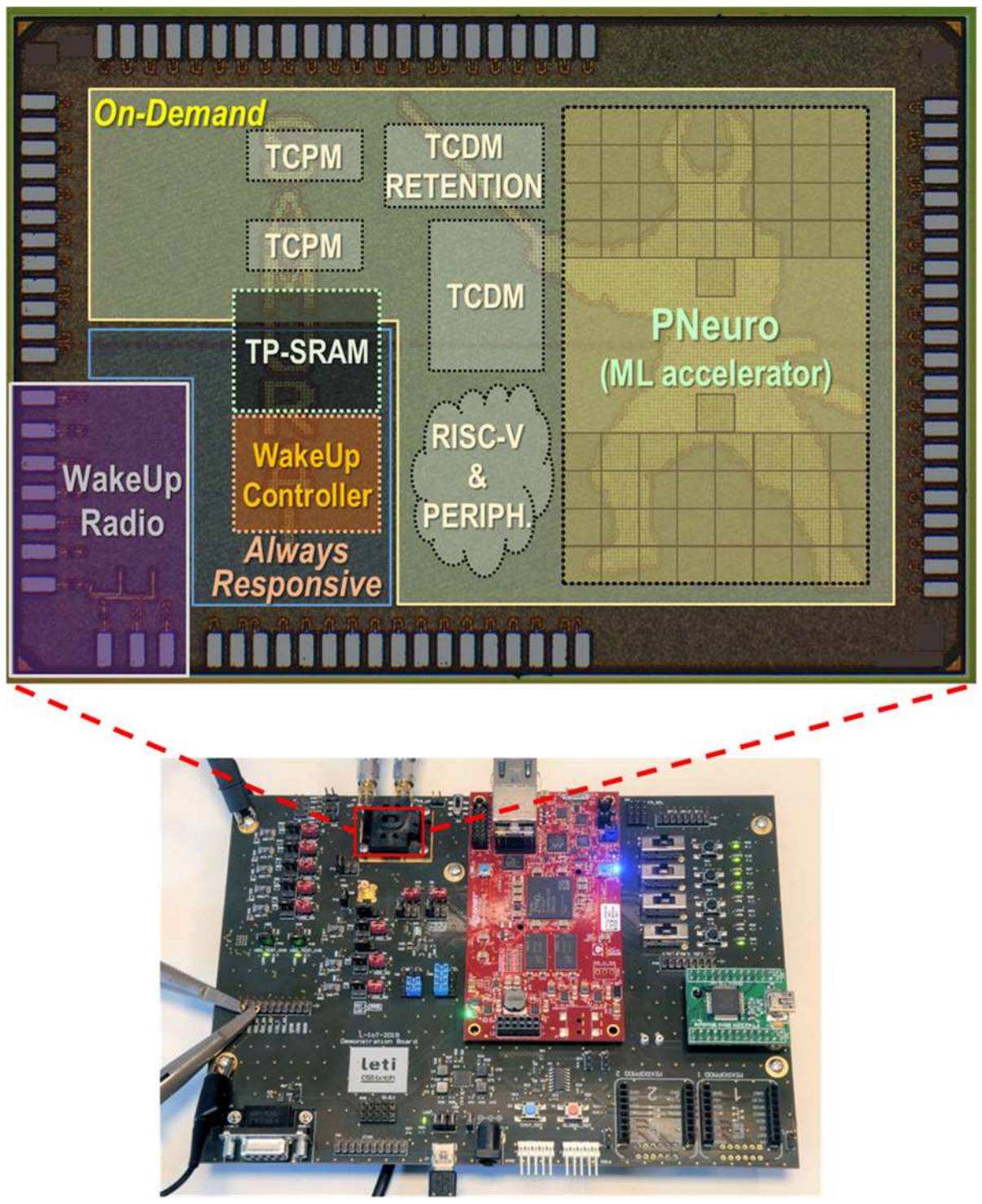

Fig. 11. SamurAI die micrograph and its test board.

## VI. SamurAI Silicon Results

SamurAI has been manufactured using ST 28nm FDSOI technology with 8 metal layers and LVT transistors. Fig. 11 shows the die micrograph and its test board. The main blocks in SamurAI are highlighted and the circuit characteristics are summarized in Table III. SamurAI also embeds testing and performance monitoring not highlighted here. The die area occupies 4.52mm². TCDM with retention and TP-SRAM memories have been designed using standard cell layout rules. Therefore, their area footprints are not optimized.

The PNeuro accelerator occupies a major portion of the circuit. Its clusters are physically placed one above the other to reduce the inter-cluster latency and congestion. A PNeuro cluster has 48 memories of 4kB for data and a 4kB memory for program. Thus, PNeuro embeds 264kB in total.

A test board (Fig. 11) was used to measure the circuit where the generated voltage sources have both voltage and current measurements. To increase the measurement precision, they can be generated by external power sources. A Zynq FPGA embedding Linux is used to execute the test scenarios on the circuit and to communicate with a PC through Ethernet.

### A. *Always-Responsive sub-system results*

The WuC controller was measured at 0.45V. Its idle power is 1.6µW while it reaches 14.45µW and 1.7MOPS when fully active. When an event arrives, the WuC's wake-up time from *IDLE* state to the first instruction fetch, takes 207ns. This is only a third of an instruction cycle time. Fig. 12 shows an annotated chronogram with the time to first fetch. These measures have been performed using a built-in performance counter, when SamurAI is in *IDLE* power mode and TP-SRAM is in sleep mode. Once an event is received, the WuC requests the wake-up of TP-SRAM in 95ns. The memory is woken-up in 15.5ns. Then the WuC performs the read operation into the TP-SRAM read port and starts the execution of the first instruction. Thanks to this fast wake-up time, the WuC and TP-SRAM have the reactivity of an instant-on system with the power consumption of the *IDLE* power mode.

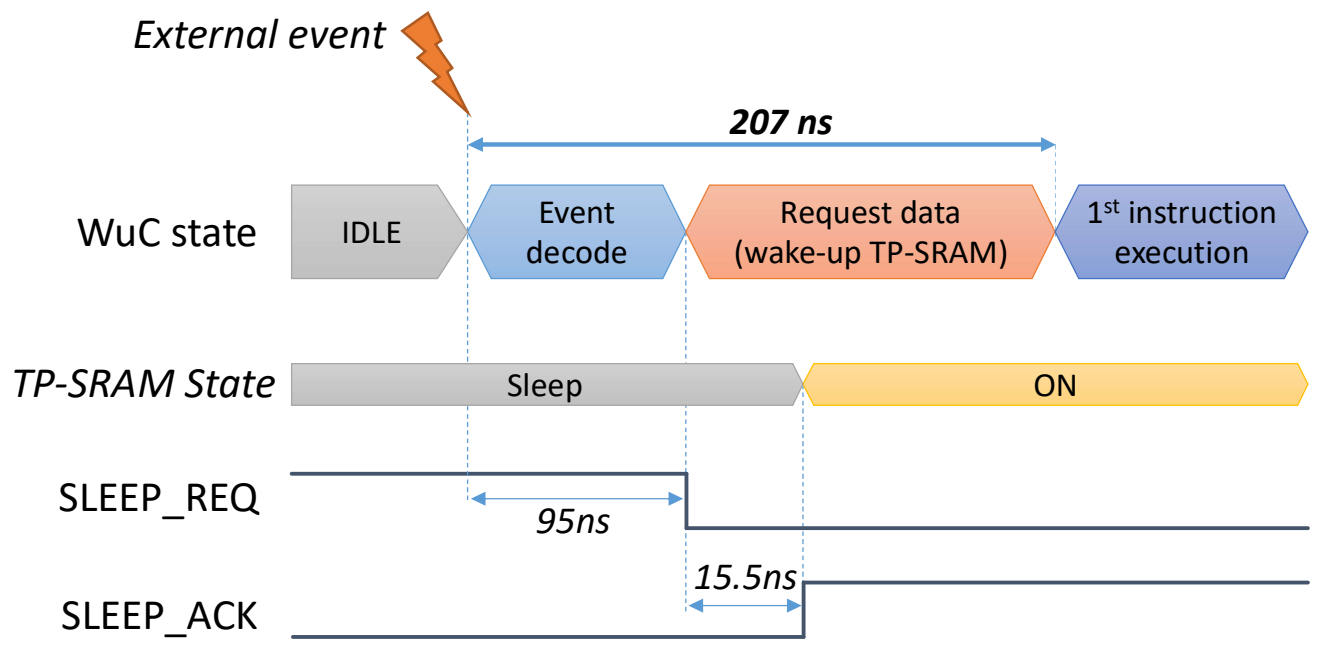

Fig. 12. Measures of wake-up time from an event to first WuC instruction fetch.

Fig. 13 shows a TP-SRAM simulation of wake-up and sleep times in function of the power supply and two process/temperature corners. The wake-up and sleep times are correlated for the simulated PVT. The measured time is in accordance with the simulated time for $V_{TP\text{-}SRAM} = 0.48V$.

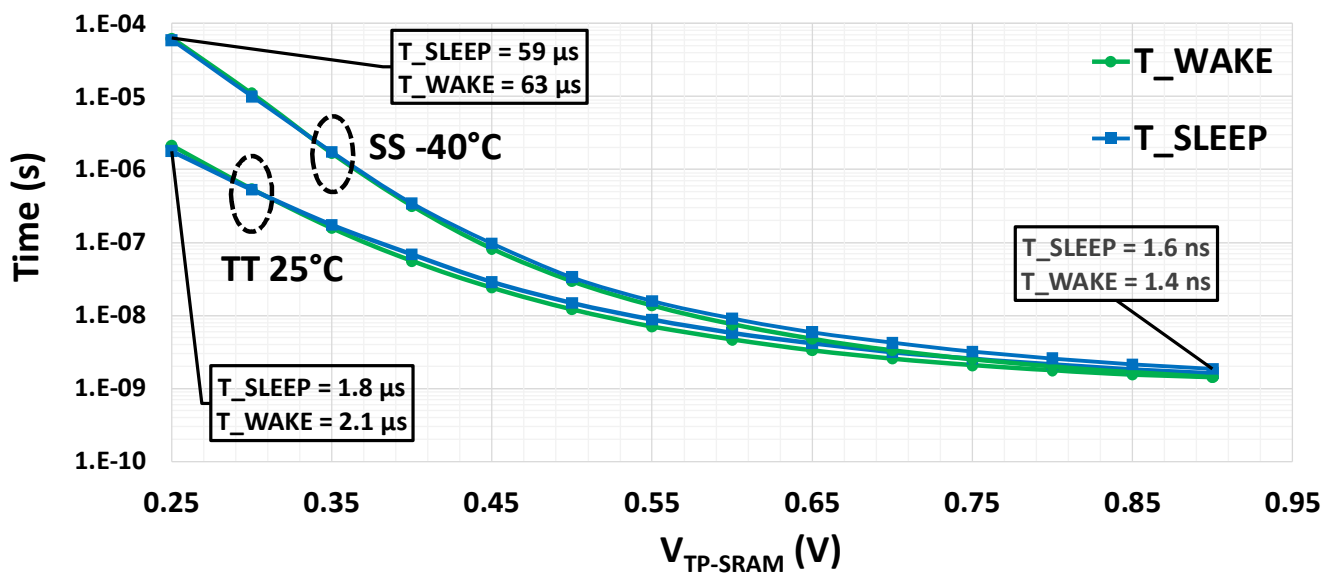

Fig. 13. TP-SRAM simulation of wake-up and sleep times in function of PVT.

TP-SRAM was measured functional down to 0.35V as shown in the shmoo plot (Fig. 14). Reading at RP and writing at WRP is possible down to 0.35V. However, reading at WRP is only possible down to 0.4V due to limited read margin. At 0.48V, its power consumption is 4.6µW in *IDLE* power mode and 14.3µW when the WuC is running at 1.7MOPS. The SVGND voltage can be tuned to achieve low operating voltages. For the ranges of 0.45V to 0.9V of main supply, this voltage can be fixed to 0.35V to simplify the voltage management.

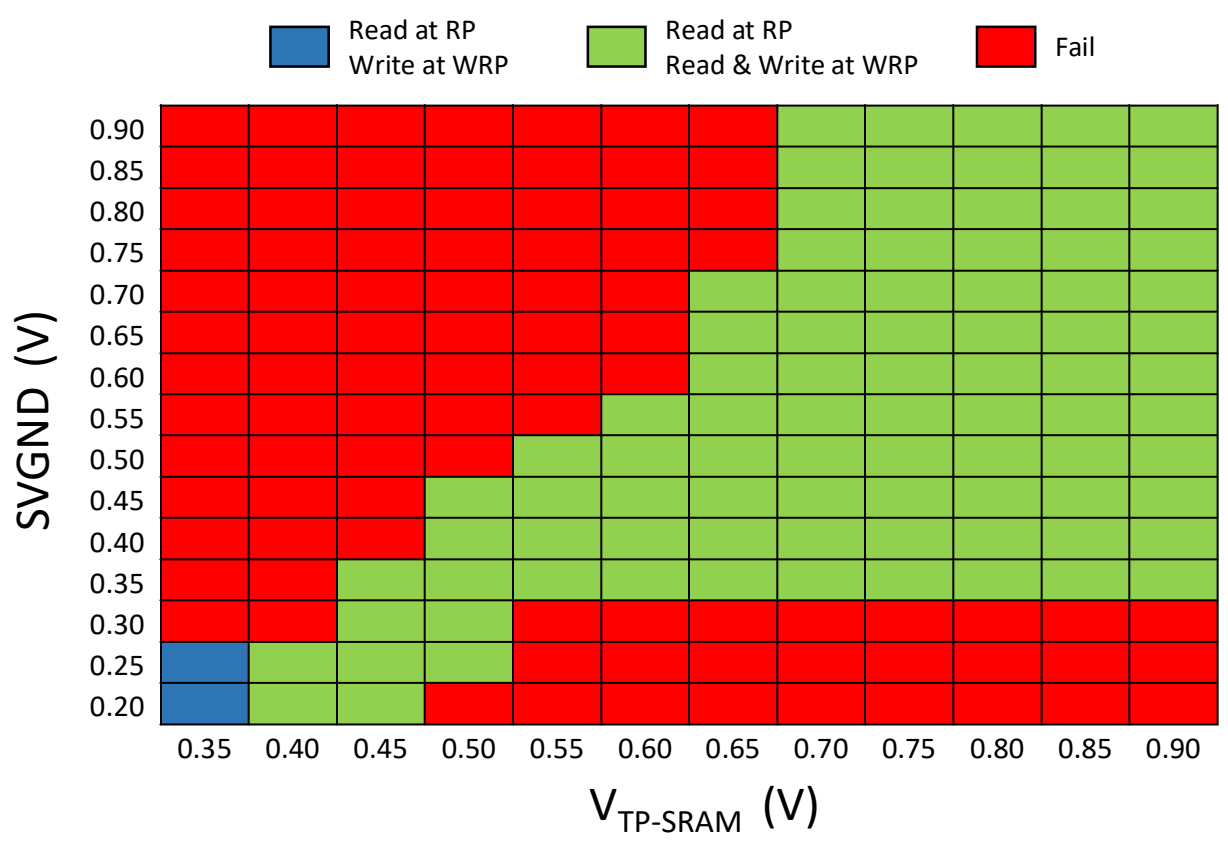

Fig. 14. TP-SRAM shmoo plot in function of its supply voltages.

To characterize the steady-state power consumption profile of the WuC with TP-SRAM, an arbitrary 2000 instructions task was executed starting and ending in IDLE power mode (Fig. 15). When the WuC is running, the profile is flat thanks to asynchronous logic (no clock spikes) and also to a run-to-completion task. Fluctuations observed in IDLE states (before and after execution) are caused by the performance monitors.

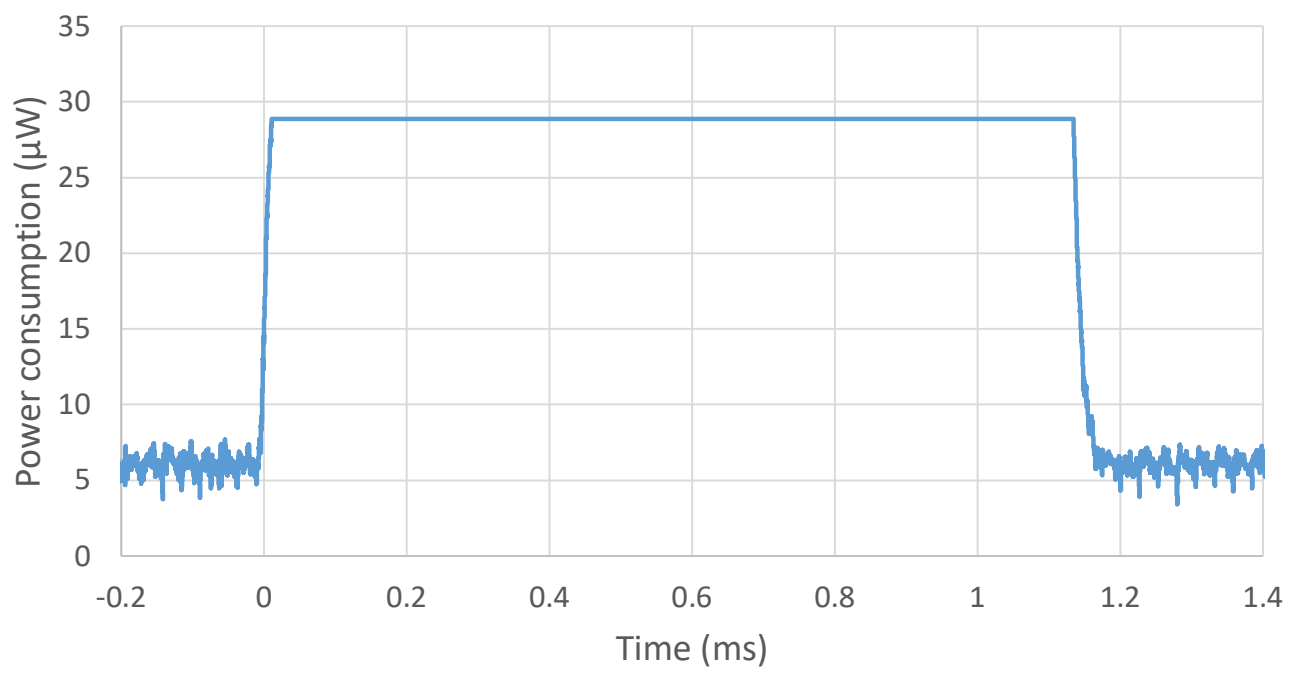

Fig. 15. Measure of WuC and TP-SRAM power consumption profile for a 2000 instruction WuC wake-up task.

The performances of the WuR have been tested on 433MHz and 868MHz bands. A function signal generator produces a pattern containing preamble and data at 50Kbps, which modulates a RF signal generator. The power is measured through the dedicated WuR 0.9V power supply. The power consumption does not exceed 76µW while decoding, and goes down to less than 4µW with a 5% duty cycle. In idle mode, it draws 40nW. The circuit was able to decode wake-up OOK patterns up to -73dBm for 433MHz and -65dBm for 868MHz.

## B. *On-Demand sub-system results*

The performances of the OD sub-system were measured in function of the RISC-V activity. The OD sub-system is functional from 0.48 to 0.9V. The measures of power consumption contain all OD sub-system elements including PNeuro, memories and peripherals as they share the same power domain. When the RISC-V is running Dhrystone and PNeuro is idle, the OD sub-system reaches 25MHz and 19pJ/cycle at 0.48V. It also reaches 350MHz and 66pJ/cycle at 0.9V, attaining a 14x higher frequency for only 3.47x higher energy per cycle.

The 32kB of TCDM with retention uses a dedicated voltage for the array. When the OD sub-system is off, it has a leakage of 1.03pA/bit at 0.5V.

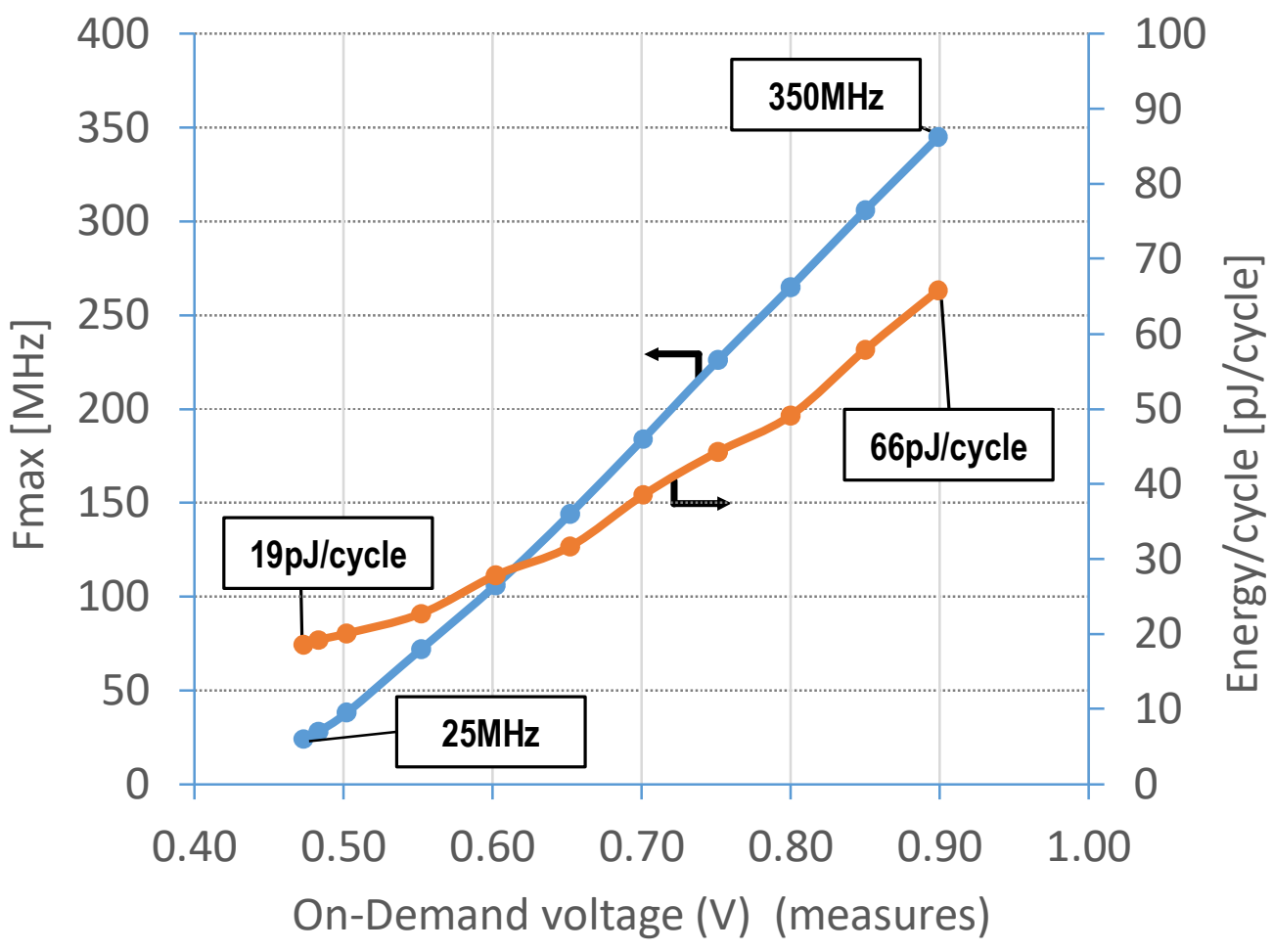

Fig. 16. Measure of Fmax and energy per cycle of OD sub-system when RISC-V is running Dhrystone and PNeuro is idle.

The PNeuro power consumption was measured when the RISC-V core is in idle state. When performing an 8-bit precision inference on a fully-connected layer, PNeuro reached 1.3TOPS/W and 2.8GOPS at 0.48V (Fig. 18). It also reaches 360GOPS/W and 36GOPS at 0.9V. Thus, it is possible to increase the network throughput by 12.8x with an energy penalty of 3.4x. For 5x5 and 3x3 convolution kernels, the efficiency reaches 1.28TOPS/W and 1.09TOPS/W respectively. When all the network weights are stored in PNeuro memories, it reaches a MAC efficiency of 89% for the fully-connected layer and 78% and 55% for 5x5 and 3x3 convolution kernels respectively.

Fig. 17 shows latency and energy results for a keyword spotting (KWS) application [44] using the depth-wise separable convolution neural network model (DS-CNN). The KWS reached 94.6% accuracy and was executed on three different configurations: two PNeuro clusters, one PNeuro cluster and only on RISC-V core. By using two PNeuro clusters instead of just one, the latency and the energy are reduced by 21% and 10% respectively. Running the application on RISC-V core would require 380x (295x) higher latency and 188x (170x) higher energy compared to two (one) PNeuro clusters.

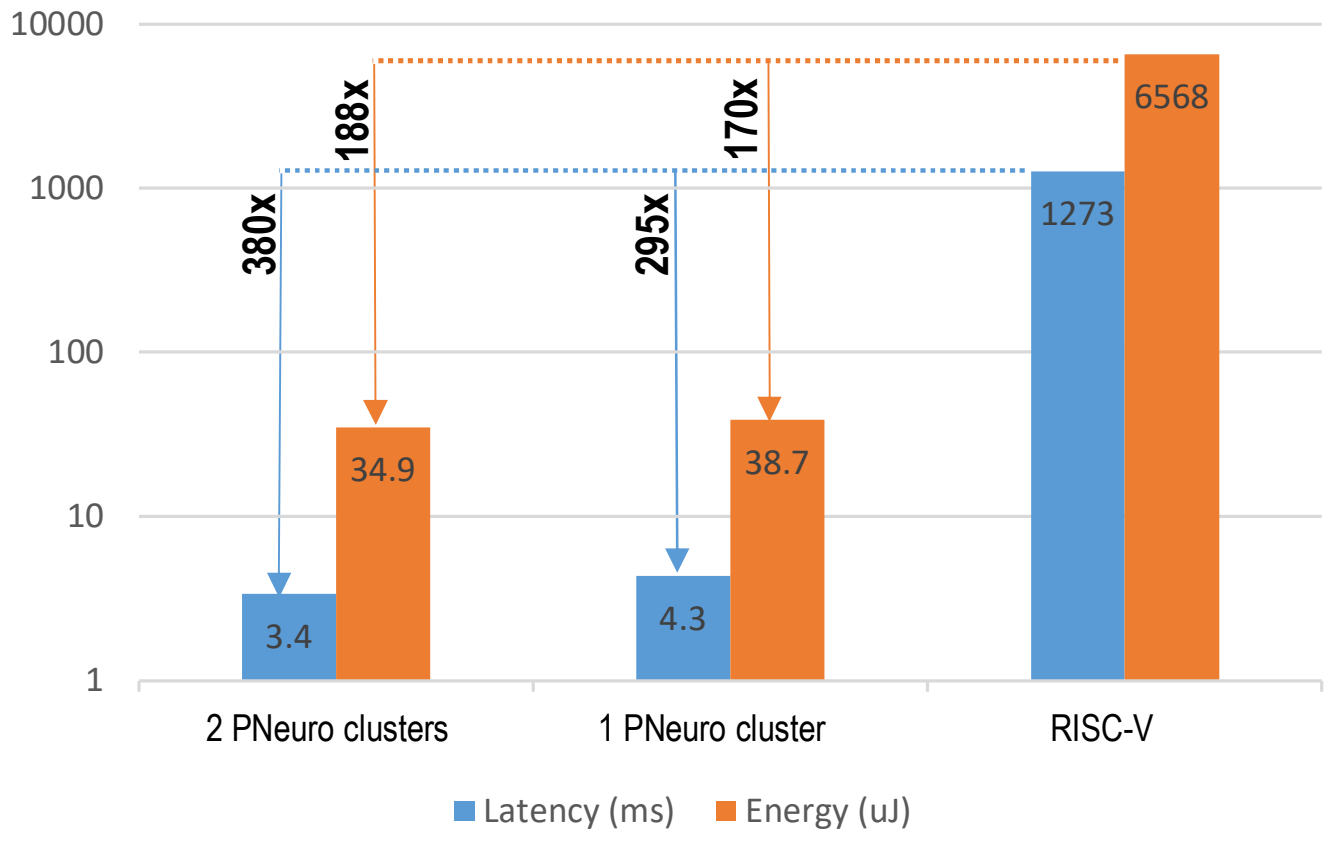

Fig. 17. Measures of KWS application [44] using 1, 2 PNeuro clusters or the RISC-V core at 100MHz.

Fig. 19a depicts the power consumption of the different power modes of the architecture, while Fig. 19b shows the power breakdown of these power modes. The *Retention SRAM* corresponds to power consumption of the 32kB of TDCM with retention. SamurAI reaches 96mW at full activity (CPU + PNeuro) and 6.4µW at IDLE mode. This 15,000x ratio between peak and idle power, highlights the adaptive and versatile performance of this dual-system architecture. Thanks to always-responsive asynchronous-logic WuC, our circuit achieves low idle power without ever entering conventional deep-sleep state: the 207ns wake-up time, which is a third of the instruction cycle, allows instant 1.7MOPS performance with reduced idle power consumption. In *IDLE* mode, the WuC and TP-SRAM power consumption are 25.1% and 72.2% respectively. In *WuC only* mode, the power consumption ratio between the WuC and TP-SRAM is similar. When the WuR and DBB are both active (*WuC+WuR* mode), the power consumption is increased by 4.1µW.

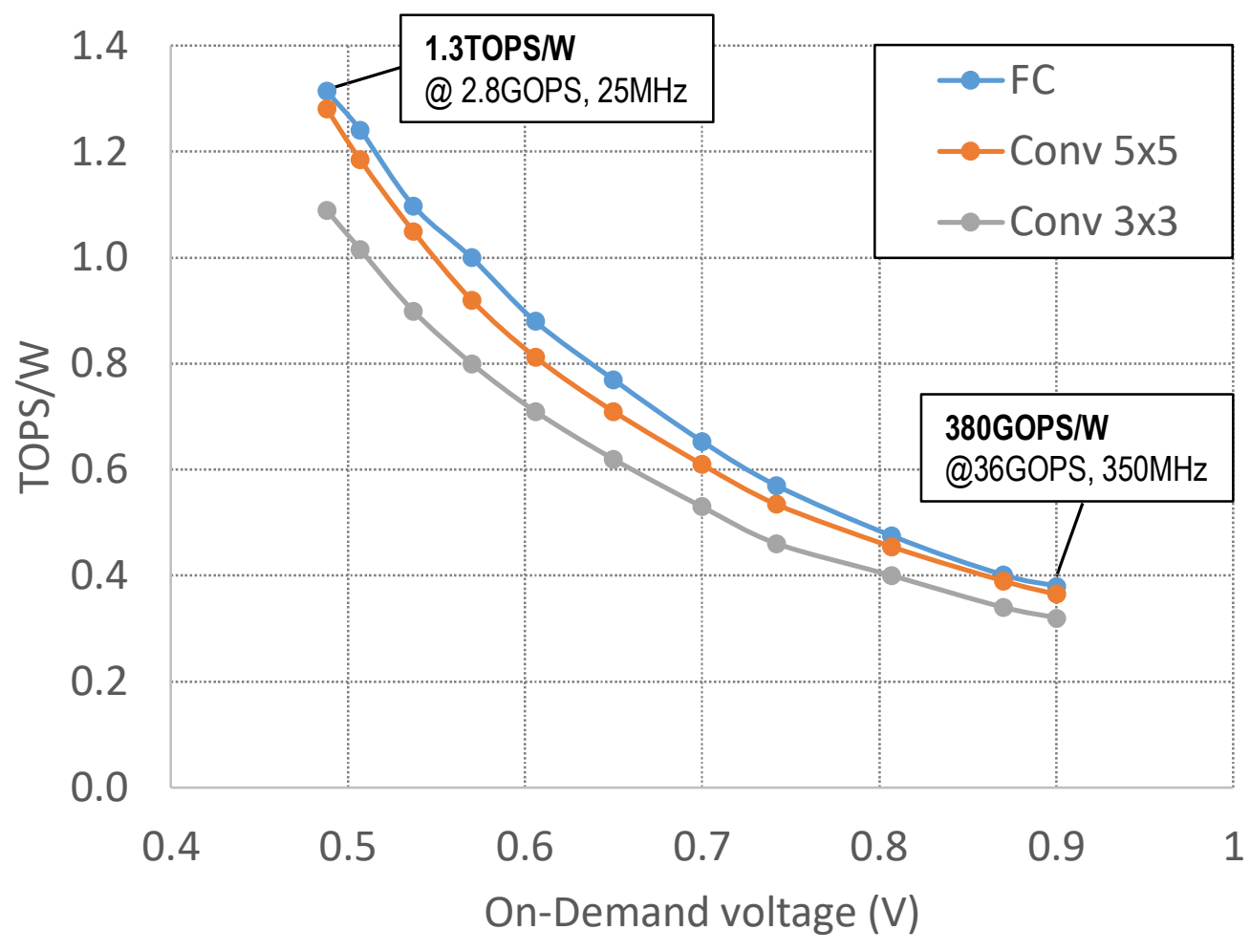

Fig. 18. Measure of PNeuro energy-efficiency and performance on different network layers types.

The WuC can wake-up the OD sub-system to interact with the OD peripherals but keeping RISC-V and PNeuro in idle mode (*WUC+Periph* mode). In this mode, the power consumption increases to 224µW where 86.6% of it is used by OD sub-system. Finally, when the circuit is running at full performance, it reaches 96mW at 0.9V and 350MHz.

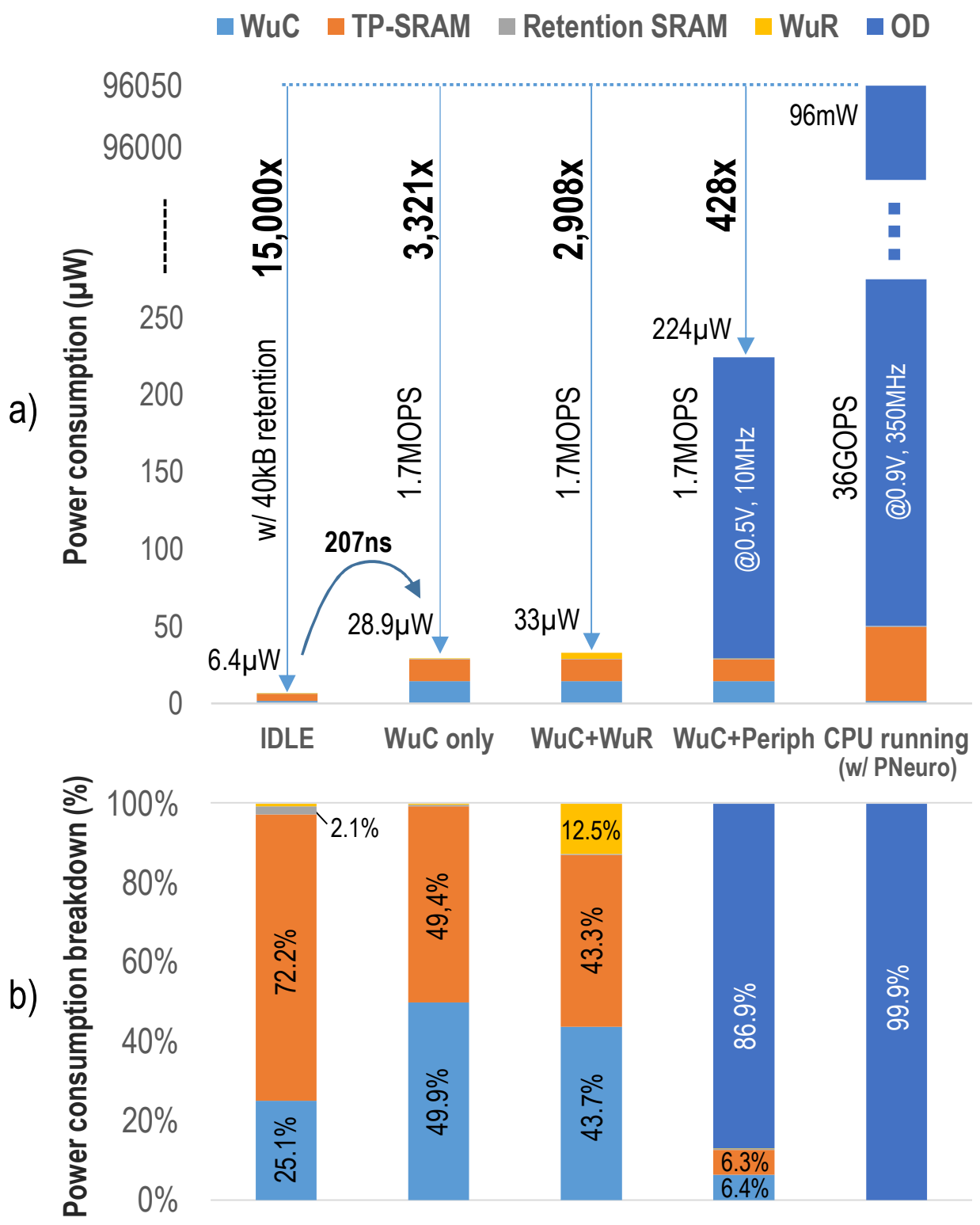

Fig. 19. Measures of power consumption, breakdown, and reduction w.r.t. the power modes.

TABLE III: SAMURAI MAIN FEATURES

| | |
|---|---|
| Technology | ST 28nm FDSOI, LVT, 8 metal layers |
| Die area | 4.52mm² |
| Voltage range | 0.45V to 0.9V |
| Frequency | Up to 350MHz |
| SRAM | 8kB TP-SRAM<br>64kB TCPM<br>128kB TCDM (32kB w/ retention)<br>264kB in PNeuro |
| Power domains | 6 |
| Deep sleep power | 6.4µW |
| Wake up time | 207ns |
| Peak performance | 36GOPS |
| Peak efficiency | 1.3TOPS/W @(2.8GOPS 0.48V) |
| WuR sensitivity | -73dBm |

Table III summarizes the main features of the circuit while Table IV compares the circuit to prior art and shows significant improvements in terms of versatility, performance, wake-up time and power reduction. None of the listed circuits has equivalent features compared to this circuit. The amount of embedded memory spans from 8kB to 580kB. A more recent design [49], achieves similar energy and performances while using programmable RI5CY cores instead of a dedicated accelerator. When compared to designs with similar retention SRAM, our design achieves the highest peak-to-idle power reduction. A proposed aggregated versatility FOM (peak performance * retention capability / idle power) shows 2x better performance for our circuit.

TABLE IV: SILICON BENCHMARK RESULTS W.R.T. STATE-OF-THE-ART

| | **VLSI 2020 This work [27]** | **JSSC 2018 Lallement [45]** | **JSSC 2018 Yu Pu [46]** | **JSSC 2017 S. Paul [11]** | **JSSCC 2016 J. Myers [5]** | **ISSCC 2017 S. Bang [47]** | **JSSC 2019 A. Pullini [48]** | **ISSCC 2021 D. Rossi [49]** |
|---|---|---|---|---|---|---|---|---|
| Technology | 28nm FDSOI | 28nm FDSOI | 28nm LP | 14nm FinFET | 65nm LP | 40nm | 40nm LP | 22nm FDSOI |
| CPU | 32b Async RISC<br>32b RISC-V<br>(RV32IMCXpulp) | M0+ | M0 | x86 IA | M0+ | M0 | 5 x RI5CY<br>RVC32IMFX | 10 x RI5CY<br>RVC32IMFX |
| Memory | 464kB SRAM | 8kB SRAM | - | 72kB SRAM<br>8kB I$<br>16kB ROM | 24kB SRAM | 270kB SRAM | 512kB SRAM<br>4kB I$<br>64kB D$ | 1.7MB SRAM<br>4MB MRAM |
| Wake-up unit | Yes | No | No | No | No | No | No | Yes |
| Wake-up Radio | Yes | No | No | No | No | No | No | No |
| ML accelerator | Yes | No | Yes | No | No | Yes | Yes | Yes |
| AVS | Yes | No | Yes | Yes | Yes | No | Yes | No |
| Crypto IPs | Yes | No | No | No | Yes | No | No | No |
| CPU state retention in deep sleep | Yes | No | - | Yes (S1) | Yes | No | Yes | Yes |
| Voltage range | 0.45V – 0.9V | 0.47V – 0.65V | 0.55 | 0.308V – 1V | 0.25V – 1.2V | 0.63V – 0.9V | 0.8V – 1.1V | 0.5V – 0.8V |
| Maximum frequency | **350MHz** | 150MHz | 50MHz | 297MHz | 66MHz | 19.3MHz | 450MHz | 450MHz |
| Deep sleep power (retention memory) | 6.4µW<br>(40kB SRAM) | 0.704µW<br>(8kB SRAM) | 1.71µW<br>(NA) | - | 80nW<br>(8kB SRAM) | - | 108µW<br>(448kB SRAM) | 1.7µW (no ret)<br>4.5µW<br>(16kB SRAM) |
| Wake-up time from deep-sleep | **207ns**<br>(35% of inst. cycle) | ~µs | - | > 1ms (S1)<br>> 1s (S0) | ~µs | - | - | - |
| GOPS | **1.7MOPS – 36GOPS** | 150MOPS | - | - | 66MOPS | - | 7GOPS | 32.2GOPS |
| Best performance | **1.3TOPS/W<br>@ 2.8GOPS** | 370GOPS/W<br>@ 16MOPS | - | 58GOPS/W<br>@ 3.5MOPS | 85GOPS/W<br>@ 750kOPS | 374GOPS/W<br>@ 107MOPS | 120GOPS/W<br>@ 2.2GOPS | 1.3TOPS/W<br>@ 15.6GOPS |
| FOM1: Peak-to-idle power reduction | **15,000x**<br>(w/ 40kB SRAM) | 51.5x<br>(w/ 8kB SRAM) | - | 4.7x | 6,940x<br>(w/ 8kB SRAM) | - | 1,416x | 29,058x (no ret)<br>10,977x<br>(w/ 18kB SRAM) |
| FOM2: Peak-performance to Idle power ratio (GOPS/µW_idle) | 5.63<br>GOPS/µW | 0.21<br>GOPS/µW | | | 0.83<br>GOPS/µW | | 0.064<br>GOPS/µW | 7.16<br>GOPS/µW |
| FOM3: FOM2 with retention capacity (GOPS*kB ret/µW_idle) | **225**<br>GOPS*kB/µW | 1.70<br>GOPS*kB/µW | | | 6.60<br>GOPS*kB/µW | | 29.04<br>GOPS*kB/µW | 114.49<br>GOPS*kB/µW |

## C. *Application scenario*

To illustrate the versatility of the architecture, SamurAI has been used to evaluate the power consumption of a smart building scenario where the presence and activity of people are analyzed. Fig. 20 shows the system view where off-the-shelf components are connected to SamurAI. Table V lists the specification for this scenario.

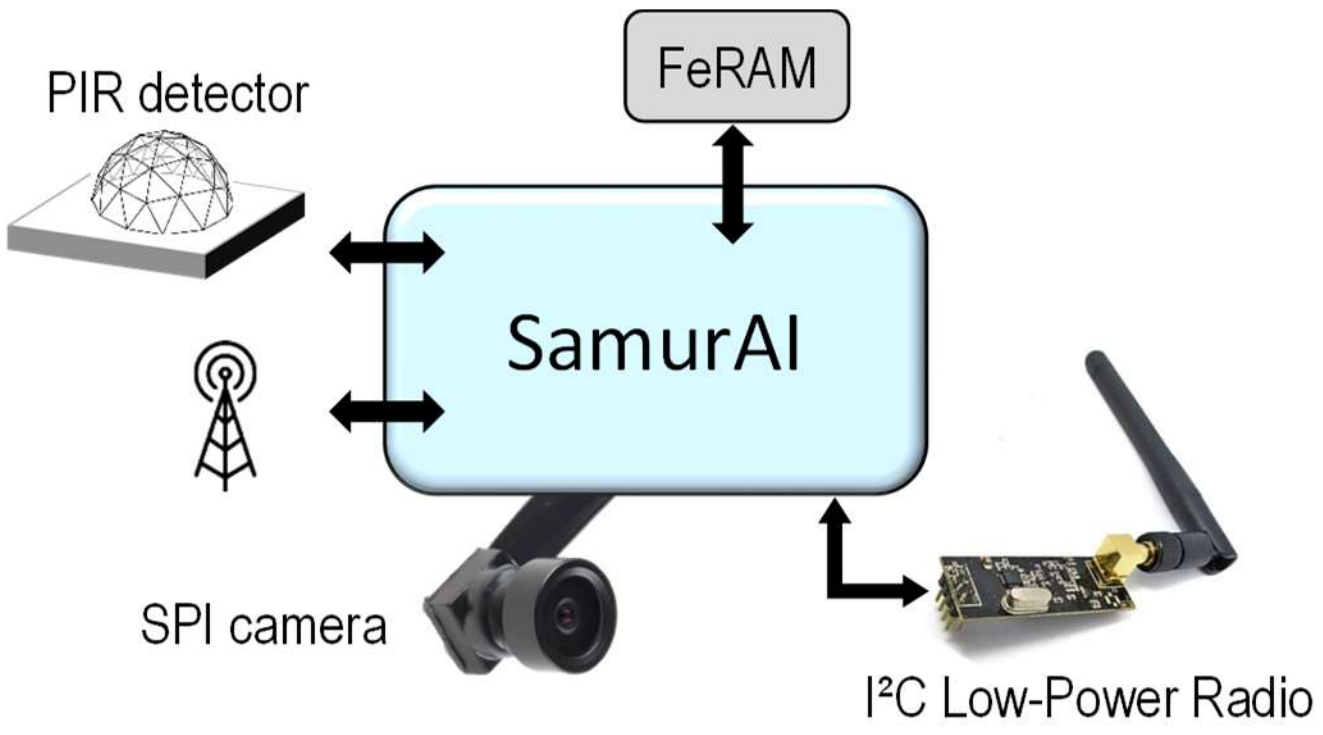

Fig. 20. System view of presence classification scenario using SamurAI with off-the-shelf components.

A pyroelectric detector (PIR) is used to detect the presence of people in a room. This kind of detector has a low power consumption and triggers each time a hot object is moving. Thus, capturing an image and analyzing it each time the PIR triggers would drain the battery of the system too quickly. To minimize power consumption, the PIR is connected to a GPIO of the WuC, waking it up at each detection. The role of the WuC is to filter PIR activity based on previous scenes classification and detection interval. Thus, it powers up the OD part only when required. When the OD part is woken up, the RISC-V acquires an image from the SPI camera and, in parallel, loads the program and the PNeuro weights from the FeRAM. Once the image is available, PNeuro is used to classify the image on a DNN with a complexity of ~100MOPS. The classification results are shared with the WuC through the TP-SRAM. WuC uses these results to manage filtering parameters of the PIR detections in function of the classification results and the time interval of PIR detections. Moreover, the RISC-V can send radio messages through an external low-power radio system. The message is first AES encrypted with the *Crypto IP* before being sent. Finally, the WuR is also used to receive user commands to configure the system.

TABLE V: SPECIFICATIONS OF PRESENCE CLASSIFICATION SCENARIO

| | |
|---|---|
| Room occupation | 8H/day |
| PIR detection interval | 5s |
| PIR power consumption | 6µW |
| Camera power consumption | 2.5mW @1FPS |
| Image size for classification | 224x224 B&W pixels |
| DNN complexity | ~100MOPS |
| External radio power consumption | 180mJ/message |
| Mean radio messages per day | 5 |

The daily average power for this application is 105µW where 26% is consumed on SamurAI (Fig. 21). The camera is the major contributor (47%) of the power consumption. Thus, reducing the number of captured and classified images is fundamental for this kind of scenarios. Under this configuration, the WuC has filtered 70% of PIR detections, reducing the total power consumption by a 2.8x factor. Furthermore, filtering 2x less PIR detections increases the power by 1.90x showing that 89% of the daily power is proportional to the filtering rate.

Moreover, the image classification on PNeuro accounts for only 1% of the total power consumption. Using RISC-V instead of PNeuro to compute the DNN would increase the total daily average power consumption by a 2.3x factor (244µW).

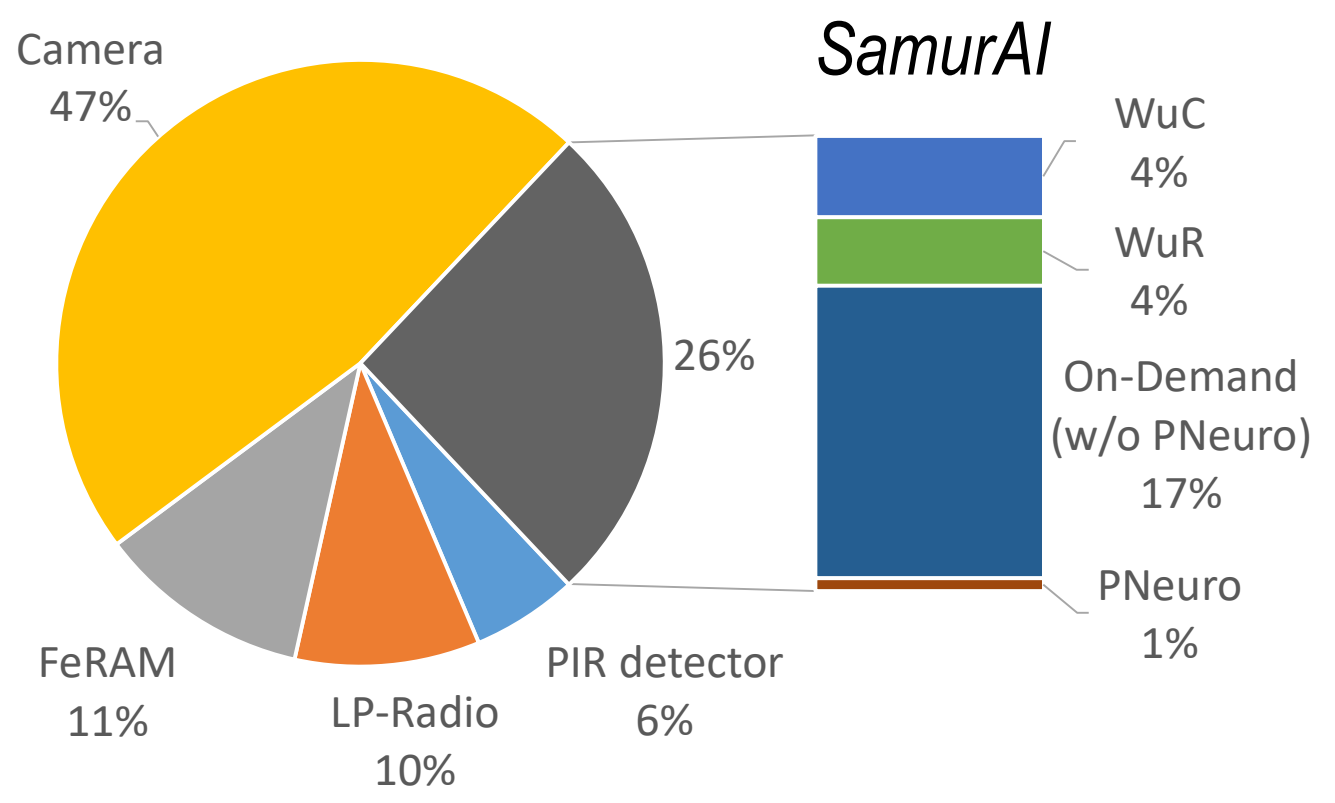

Fig. 21. Daily average power breakdown of presence classification scenario (70% PIR filtering), 105µW total power.

On a cloud-based processing where every image is transferred to the cloud to be classified and assuming 3.5nJ/b BLE link [50], the mean power of the scenario (without considering the cloud power consumption, only the node) would be higher by a 3.5x factor (366µW). 25.8% of the power is spend on the radio link and 45.6% on the camera. These gains demonstrate the relevance of adaptive event-based filtering in AR part and local ML processing in OD part of SamurAI.

## VII. Conclusion

The presented SamurAI circuit fills the gap between sporadic computing and energy-hungry applications on IoT domain by leveraging two on-chip sub-systems. An asynchronous WuC optimized for short sporadic computing with a 207ns wake-up time and a generic RISC-V core coupled with an energy-efficient ML accelerator are the computing pillars to achieve this versatile architecture. The node can perform up to 36GOPS and achieves 1.3TOPS/W peak energy efficiency. Thanks to the proposed AR/OD approach, 15,000x reduction from peak-to-idle power consumption can be achieved. This approach performs efficient information filtering from sensor to radio communication, starting from handling of OD wake-up events within the asynchronous wake-up controller, then thanks to embedded image classification in the ML accelerator, allowing for up to 3.5x extension of battery lifetime on an application scenario.

## Acknowledgment

The authors would like to thank STMicroelectronics for the circuit manufacturing and for the support on retention SRAM and TP-SRAM design, PULP project for the RI5CY core, I2C and UART peripherals, Anthony Quelen for the LDO design, Reda Boumchedda for the TP-SRAM design, Edith Beigné for the L-IoT approach, Mounir Abaline and Thomas Mesquida for developing the test environment, David Turgis for TP-SRAM support, Nirmal Pregassame for the UPF flow, and Mikael Le Coadou for the FRAM controller. The authors would like to thank many other contributors from CEA who made this circuit a successful realization.

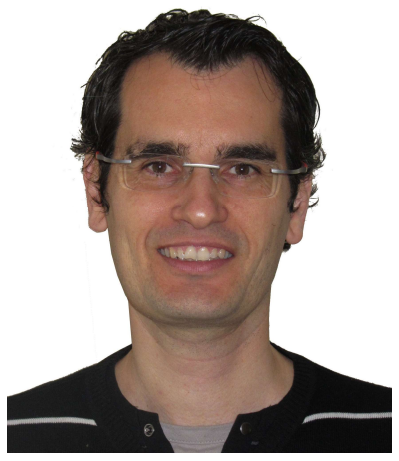

**Ivan Miro-Panades** (M'17) received a M.S. in telecommunication engineering from the Technical University of Catalonia (UPC, Barcelona, Spain) in 2002, a M.S. and his Ph.D. in computer science from the University Pierre & Marie Curie (UPMC, Paris, France) in 2004 and 2008, respectively. Dr. Miro-Panades worked at Philips Research and ST Microelectronics before joining in 2008 CEA, Grenoble, France, where he is currently an expert research engineer in digital integrated circuits. His main research interests are Artificial Intelligence, Internet-of-Things, low-power architectures, energy-efficient systems, and Fmax/Vmin tracking methodologies.

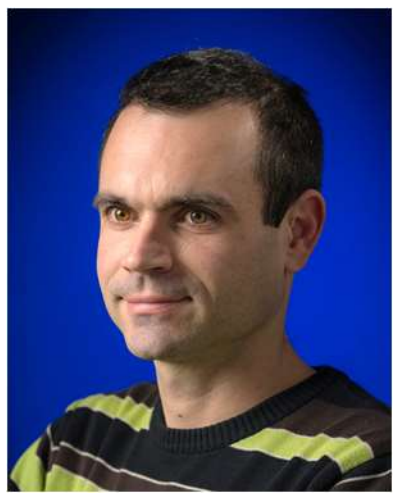

**Benoit Tain** received the M.S. degree in Electrical Engineering, from the Polytech Institute of Clermont-Ferrand, France, in 2003. Since December 2003, he is a research engineer in CEA LIST, where he works mainly in the field of embedded system. His research interests include SOC and multi-core SOC architecture and digital design for many applications including neural networks, multimedia or system vision. He has worked on several European projects, and now he is mainly involved in industrial research programs.

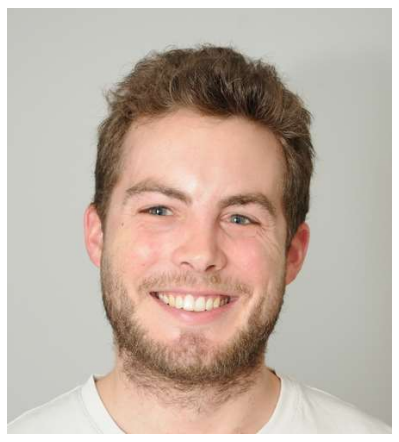

**Jean-Frédéric Christmann** (M'14) received the M.S and PhD degrees in Electronics from the Grenoble Institute of Technology in France in 2009 and 2013. Since 2009, he is a research engineer at the French Atomic Energy Commission (CEA) with the Digital Circuits and Systems Department. His fields of expertise include low power digital and mixed-signal electronics, asynchronous logic, In-memory computing, power management and energy harvesting. He is the author or co-author of more than 30 international conference papers and 10 journal papers. His research interests tackle low power edge computing, advanced design automation and quantum computing.

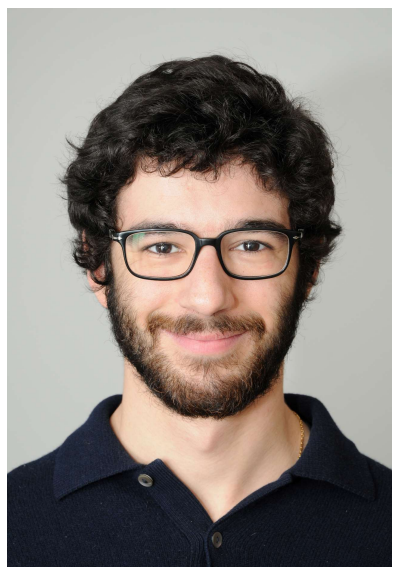

**David Coriat** received his M.Sc. degree for the University of Science of Montpellier, France, in 2012, and subsequently joined the CEA-LETI Minatec. He has worked on dynamic management of power and variability in MP-SoC architectures as well as power estimation techniques in large MP-SoC architectures. His research interests now lie in low power architectures and design.

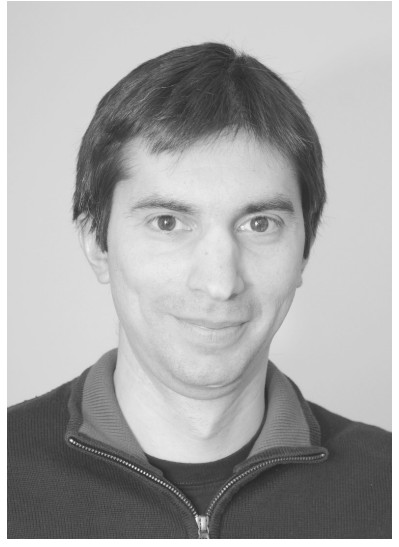

**Romain Lemaire** received the engineer degree in electrical engineering from Supelec in 2002 and the Ph.D. degree in microelectronics from Grenoble Institute of Technology in 2006. He joined CEA-Leti, as a research engineer in digital design and system architecture, in 2006. He has contributed to the design of advanced on-chip communication and control architectures for heterogeneous system-on-chips. Since 2020, he is working for CEA-List on architecture and design for embedded AI.

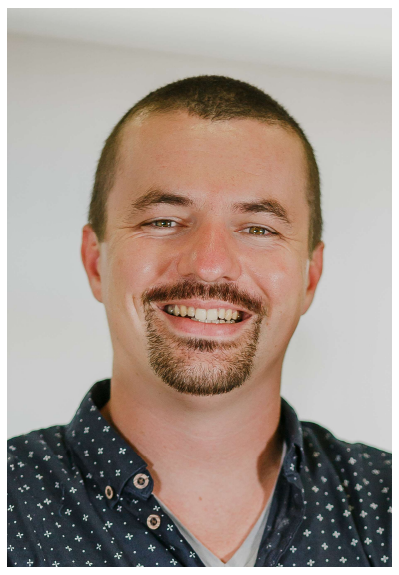

**Clément Jany** received the M.S. degree from the ENSEA, Cergy, France, in 2011. He received the Ph.D. degree from the University of Grenoble, France, in 2014. During his doctoral studies, he worked on the design and modeling of CMOS mmW oscillators and he proposed a new architecture for mmW frequency synthesis. In 2014, he joined CEA-LETI in Grenoble, France, as a research engineer. His work focused on the design of RF and mmW CMOS transceivers. Dr Jany is the recipient of the 2014 Microwave Award of the European Microwave Association, and he holds several patents. He spent the year 2018 as visiting scholar within Pr Amin Arbabian group at Stanford University working on mmW CMOS design for Tb/s datarates and for Healthcare applications. Since March 2020, he serves as CTO of TiHive, a french company leader in THz AI systems for Non Destructive Inspection.

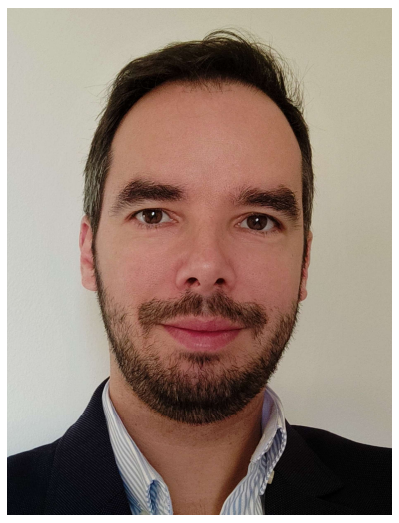

**Baudouin Martineau** received the M.S. degree in microelectronics from the Sciences University of Bordeaux 1, France, in 2004 and the Ph.D. degree in microwave and microtechnology from the University of Lille, France, in 2008. He did his Ph.D. work at the Institute of Electronics, Microelectronics and Nanotechnologies (IEMN), Villeneuve d'Ascq, France and STMicroelectronics, Crolles, France. His Ph.D. thesis focused on the 65nm CMOS SOI potentialities for millimeter wave wireless applications. In 2008, he joined the Technology R&D department of STMicroelectronics, Crolles, France, as millimeter wave and RF design engineer. Since 2014, he is with CEA LETI research Institute, Grenoble, France as expert in RF & mmW design. He has authored and co-authored more than 50 IEEE publications and more than 20 patents.

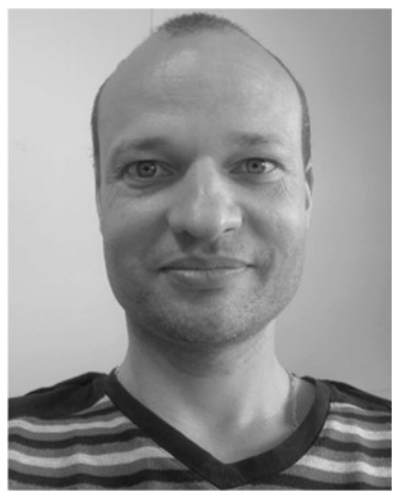

**Fabrice Chaix** was born in France in 1974. He received the M.S. degree in microelectronics from the Science University of Grenoble, Grenoble, France, in 2000. In 2001, he joined CIRIEL, Grenoble, France, where he worked on various research projects as an Analog Circuit Designer. In 2004, he joined EASII-IC, Grenoble, France, where he was involved with high-speed ADC circuit in design CMOS advanced technologies. In 2008, he joined CEA-LETI, Grenoble, France, where he is involved with analog and millimeter-wave circuit design in CMOS advanced technologies

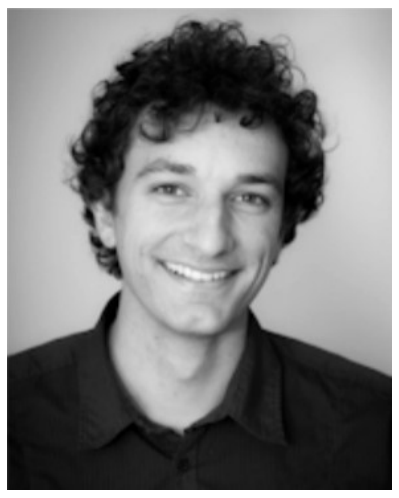

**Guillaume Waltener** received the Engineering Diploma degree from the École nationale supérieure d'électronique, informatique, télécommunications, mathématique et mécanique de Bordeaux, Bordeaux, France, with a specialization in analog design, in 2012. He was with Thales Avionics, Bordeaux, in 2011. He joined the CEA-LETI, Grenoble, in 2012, where he was involved in silicon photonic links, millimeter-wave ICs, and signal integrity in high speed interconnects. He also worked on wake-up radios, ultra-low power devices, and analog baseband circuits with industrial constraints.

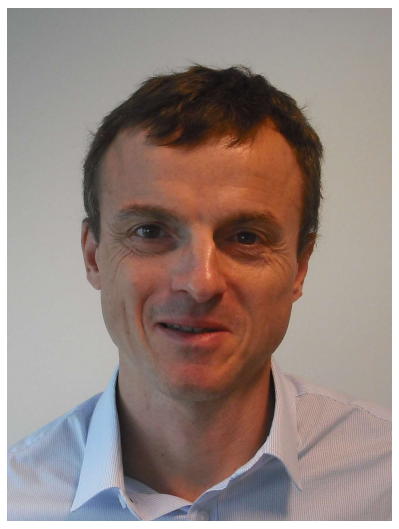

**Emmanuel Pluchart** received the M.S. degree in material sciences from PolyTech Grenoble in 1994 and the M.S. degree in microelectronics from Grenoble INP Phelma, in 1998. From 1998 to 2017, he worked in various EDA companies as an application engineer. Since 2017, he has been a research engineer at CEA LETI then CEA LIST, Grenoble.

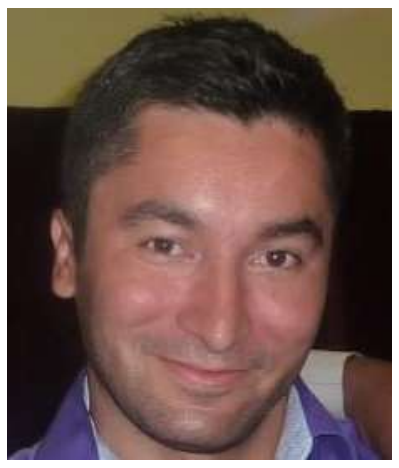

**Jean-Philippe Noel** was born in Alès, France, in 1985. He received the M.S. and Ph.D. degrees in microelectronics from Polytech Marseille, France, and the University Grenoble Alpes, France, in 2008 and 2011, respectively. During his Ph.D. studies, he worked at the CEA, Grenoble, France, on the design optimization of low power digital and memory circuits in UTBB FD-SOI technology. In November 2011, he joined STMicroelectronics as a memory design engineer. His work was focused on the design of innovative SRAM and TCAM circuits in UTBB FD-SOI technology. Since December 2016, he has joined CEA to work on innovative in- and near-memory computing architectures and emerging non-volatile memories (ReRAM, PCM …) circuit design. He has authored and co-authored over 60 publications and 30 patents.

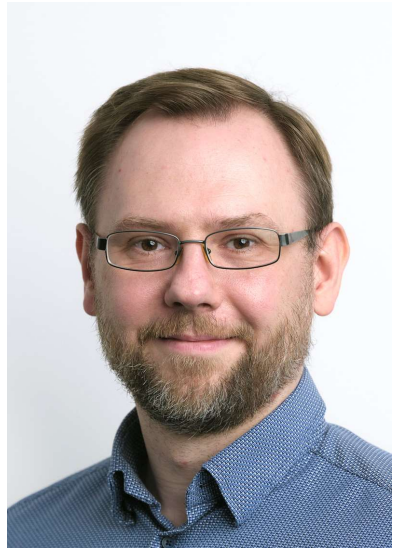

**Adam Makosiej** received the B.S. and M.S. degrees in electronics and telecommunication from the Technical University of Lodz, Poland, in 2008 and the Ph.D. degree in electrical engineering from Telecom Paristech, Paris, France, in 2012. From 2013 to 2019, he was a Research Engineer with the CEA-LETI, Grenoble, France. Since 2019, he has been working as a Product Manager in Xenergic, Lund, Sweden.

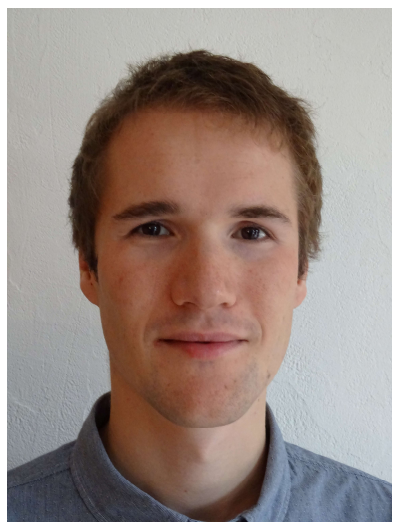

**Maxime Montoya** received both his Master of Science in Engineering from Ecole Centrale de Lyon and his Master of Science in Electronics from Université Claude Bernard Lyon 1 in 2016. He received his Ph.D. degree in Microelectronics from Mines Saint-Etienne in 2019. He is now working as Security Engineer at NXP Semiconductors Austria. His main research focus is energy-efficient hardware security for the Internet of Things.

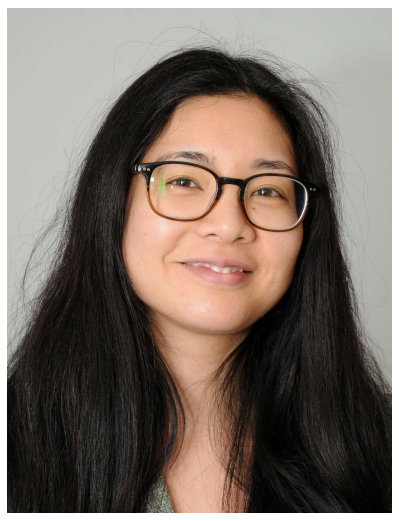

**Simone Bacles-Min** received the M.S. degree in microelectronics from the Grenoble Polytechnical Institute, Grenoble, France, in 2006. In 2014, she joined CEA, Grenoble. Her current research interests include hardware–software codesign, low-power, adaptive security circuit techniques and eco-design of ICT.

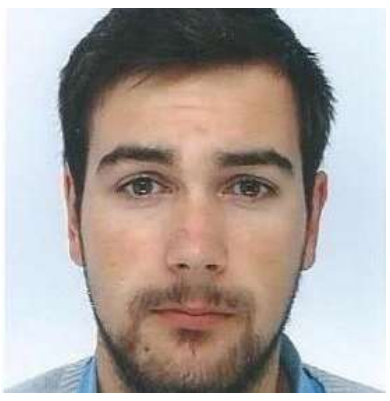

**David Briand** received a Master degree of Engineering in Electronics and Computer Science from INSA Rennes in 2014. He joined the CEA List in 2015 where he is a research engineer specialized in embedding deep learning algorithms on low power devices. He contributes to the N2D2 deep learning framework with NVidia TensorRT export module, quantization aware training, mixed low-bits quantized kernels for ARM and the first version of PNeuro code generator. David research interests focuses on network optimization with high quantization.

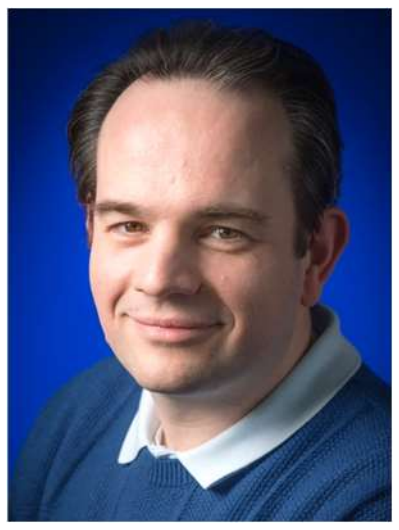

**Jean-Marc Philippe** received an engineering degree from Ecole Internationale des Sciences du Traitement de l'Information (EISTI) and a post-graduate diploma (Diplôme d'Etudes Approfondies) in signal and image processing from University of Cergy-Pontoise (France) in 2002. He also received a PhD degree in computer engineering from the University of Rennes I, France, in 2005. From 2005 to 2021, he was a research engineer at CEA, working in embedded computing and AI accelerator architectures. Since 2021, he has been a research engineer in edge computing hardware architectures at Thales Research & Technology, France.

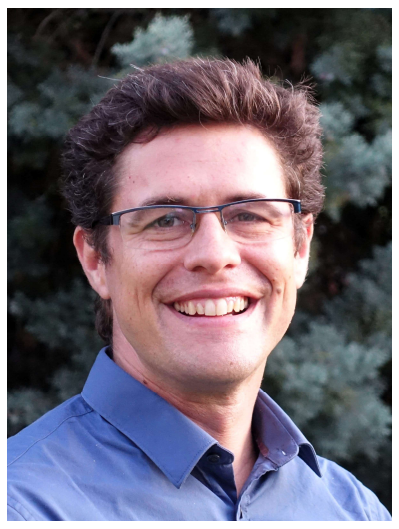

**Yvain Thonnart** (M'16) received the MS degree from Ecole Polytechnique and an engineering diploma from Telecom Paris, France in 2005. He then joined the Technological Research Division of CEA, the French Alternative Energies and Atomic Energy Commission, within the CEA-Leti institute until 2019, then within the CEA-List institute. He is now senior expert on communication and synchronization in systems on chip, and scientific advisor for the mixed-signal design lab. His main research interests include asynchronous logic, networks on chip, physical implementation, emerging technologies integration such as photonics, cryoelectronics and interposers.

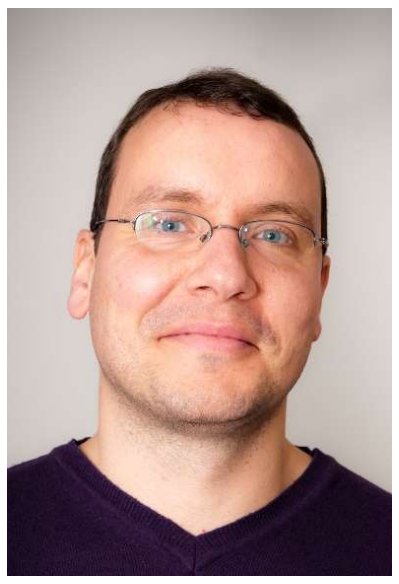

**Alexandre Valentian** after an MSc and a PhD in microelectronics, Alexandre Valentian joined CEA in 2005. His past research activities included design technology co-optimization, promoting the FDSOI technology (notably through his participation in the SOI Academy), 2.5D/3D integration technologies and non-volatile memory technology. He is currently pursuing the development of bio-inspired circuits for AI, combining memory technology, information encoding and dedicated learning methods. Since 2020, he heads the Systems-on-Chip and Advanced Technologies (LSTA) laboratory. Dr Valentian has authored or co-authored 80 conference and journal papers.

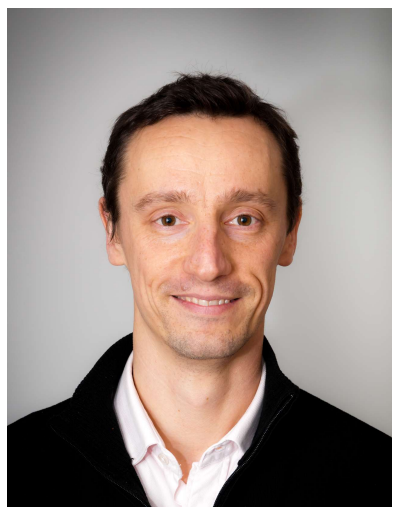

**Frédéric Heitzmann** received a master degree from the école polytechnique (France) in 2000, and a master degree from Telecom Paritech 2002. From 2002 to 2007, he worked in Multimedia and Mobile Phone Companies. Since 2007, he joined CEA-LETI and focused on hardware-software-algorithms co-development. He started the Embedded AI strategic program in 2019. Since 2020, he has been Program Manager in the Sensors and Electronic for Energy Department, in the System Division.

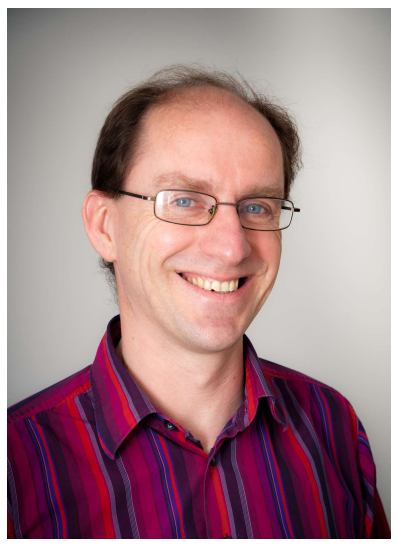

**Fabien Clermidy** is currently heading the digital architecture and IC design department in the research technology division of CEA. In this position, he develops the hardware and software computing strategy for High-Performance Computing, Artificial Intelligence, Cybersecurity and Cyber-Physical-Systems in relation with system developments in automotive, factory of future, avionics or defense and new nanotechnologies such as 3D chip-stacking, embedded Non-Volatile-Memories, photonic and quantum computing. Fabien has been working in CEA since 2000, holding different positions as project leader and manager of different teams. Fabien is a CEA research director and he has published more than 100 papers in some of the greatest conferences such as ISSCC or DAC. He is also author or co-author of 15 patents.